%% file: main_resub.tex
\documentclass[iop,twocolumn]{emulateapj}
\submitted{Submitted 20 Feb 2015}
\usepackage{graphicx}
\usepackage{float}
\usepackage{amsmath}

\date{\today}

\begin{document}
\title{Beasts of the Southern Wild: Discovery of nine 
  Ultra Faint satellites in the vicinity of the Magellanic Clouds.}
\author{Sergey E. Koposov}
\author{Vasily Belokurov}
\author{Gabriel Torrealba}
\author{N. Wyn Evans}
\email{koposov@ast.cam.ac.uk,vasily@ast.cam.ac.uk}
\affil{ Institute of Astronomy, Madingley Road, Cambridge CB3 0HA, UK}
\begin{abstract}
We have used the publicly released Dark Energy Survey data to hunt for
new satellites of the Milky Way in the Southern hemisphere. Our search
yielded a large number of promising candidates. In this paper, we
announce the discovery of 9 new unambiguous ultra-faint objects, whose
authenticity can be established with the DES data alone. Based on the
morphological properties, three of the new satellites are dwarf
galaxies, one of which is located at the very outskirts of the Milky
Way, at a distance of 380\,kpc. The remaining 6 objects have sizes and
luminosities comparable to the Segue~1 satellite and can not be
classified straightforwardly without follow-up spectroscopic
observations. The satellites we have discovered cluster around the LMC
and the SMC. We show that such spatial distribution is unlikely under
the assumption of isotropy, and, therefore, conclude that at least
some of the new satellites must have been associated with the
Magellanic Clouds in the past.
\end{abstract}
\section{Introduction}

Currently, there are no strong theoretical predictions as to the
luminosity function and the spatial distribution of the Galactic dwarf
companions. In other words, today's semi-analytic models are so
flexible they can easily produce any number of Milky Way (MW)
satellites: from a few tens up to several thousands
\citep[e.g.][]{koposov09}. Similarly, a large range of spatial
arrangements is possible: from nearly isotropic to strongly planar
\citep[e.g.][]{bahl14}. This is why to improve our understanding of
the physics of the Universal structure formation on small-scales we
look to observations. However, to date, only a third of the sky has
been inspected. The Sloan Digital Sky Survey (SDSS) has demonstrated
the power of deep wide-area imaging to fuel resolved stellar
populations studies of the Galactic halo. The analysis of the SDSS
data covering only 1/5 of celestial sphere has more than doubled the
number of the known Galactic dwarf satellites \citep[see
  e.g.][]{willman10,belokurov13}. Recently, both PanSTARRS and VST
ATLAS have expanded the surveyed region significantly, but found
little of the wealth they came to seek \citep{laevens14,belokurov14}.

The SDSS discoveries extended the dwarf galaxy regime to extremely low
luminosities and sizes. As a result of adding these new faint
surface-brightness objects (known as ultra-faint dwarfs, UFDs) to the
panoply of Milky Way companions, it has been revealed that there
appears to be a gap in the distribution of effective radii between
globular clusters and dwarfs which extends across a large range of
luminosities. In the absence of a working definition of a galaxy, an
ad hoc combination of morphological, kinematic and chemical properties
is required to be classified as one.  Therefore, some objects that
live dangerously close to the notional gap may risk
misclassification. These faintest of the UFDs are only detectable with
surveys like SDSS out to a small fraction of the MW virial radius. As
a result, their total number in the Galaxy must be reconstructed under
the assumption of the shape of their galactocentric radial
distribution. Given that only a handful of such objects are known
today, their radial profile is not observationally
constrained. However, if it is assumed that the faintest of the UFDs
are linked to the small-mass field dark matter (DM) sub-halos, their
distribution can be gleaned from cosmological zoom simulations. This
is an example of so-called sub-halo abundance matching which predicts
that the bulk of the Galactic dwarf galaxy population is in objects
with luminosities $M_V< -5$. The details of the semi-analytic
calculations may vary but the result remains most astounding: there
ought to be hundreds of galaxies with $M_V\sim -1$, if the faintest of
the UFDs occupy field DM sub-halos \citep[see
  e.g.][]{tollerud08,bullock10}.

It is, however, possible that the smallest of the UFDs may have been
acquired via a different route. It has been suggested that some of
these objects may have been accreted as part of a group \citep[see
  e.g.][]{belokurov13}. In the case of the UFD Segue 2 for example,
the group's central object appears to have been destroyed and can now
be detected only as tidal debris in the MW halo
\citep{belokurov09,deason14}. In this scenario, the satellites of
satellites survive (unlike their previous host) in the MW halo, but
their spatial distribution should differ dramatically from that
predicted for the accreted field sub-halos. Not only the radial
density profile is modified, the satellites' morphological properties
might have been sculpted by the pre-processing \citep[see
  e.g.][]{wetzel15}. Thus, the total numbers of the UFDs would be
biased high if estimated through simple abundance matching
calculation, while their spatial anisotropy would be greatly
under-estimated.

On the sky, the remaining uncharted territory lies beneath declination
$\delta=-30^{\circ}$ and is currently being explored by the rapidly
progressing Dark Energy Survey (DES).  This is the realm of the
Magellanic Clouds, long suspected to play a role in shaping the
Galactic satellite population.  For example, \citet{lyndenbell76}
points out the curious alignment between several of the Milky Way
dwarfs and the gaseous stream emanating from the Magellanic
Clouds. {\citet{kroupa05} developed this picture, adding more Milky Way
satellites to the Great Magellanic family. The idea also receives
support from the planar structures of satellites around M31 and other
nearby galaxies~\citep[see e.g.,][]{Ib13,Tu15}.} To explain the
peculiar orbital arrangement of the Large Magellanic Cloud (LMC) and
the fainter dwarfs within the $\Lambda$CDM paradigm,
\citet{odonghia08} consider a cosmological zoom simulation of
accretion of an LMC-type object onto a Milky Way-size galaxy. In their
simulation, approximately 1/3 of the satellites around the prototype
MW at redshift $z=0$ come from the LMC group. They conjecture that the
process of dwarf galaxy formation at the epochs around the
re-ionisation is favoured in group environments, suggesting that at
least in part, this can be explained by a significant difference in
gas cooling times in the halo of a large galaxy like MW and a small
group of the size of the LMC.

In this paper, we announce the discovery of a large number of new
faint satellites detected via an automated stellar over-density search
in the DES public release data. The paper is structured as
follows. Section 2 outlines the steps required to produce object
catalogues from the calibrated DES image frames. Section 3 presents
the details of the satellite search routine. Section 4 gives an
overview of the properties of the detected satellites, while Section 5
discusses their spatial distribution. The paper closes with Discussion
and Conclusions in Section 6.

\begin{figure}
\includegraphics[]{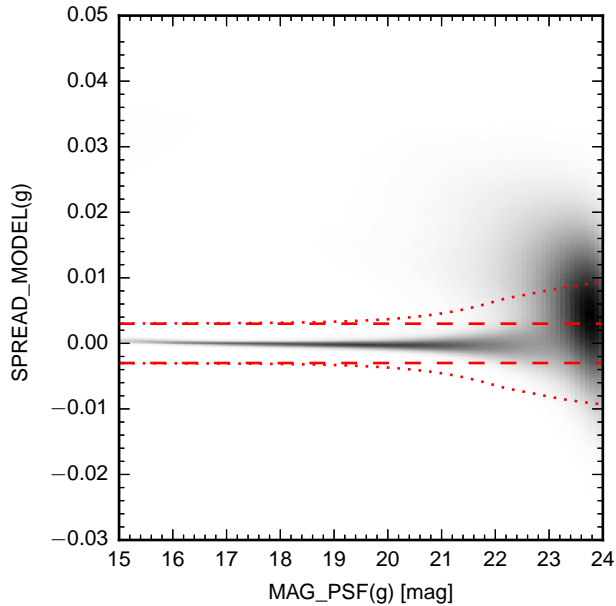}
\caption{SExtractor star/galaxy indicator SPREAD\_MODEL as a function
  of the PSF magnitude. The red dashed curves are the lines of
  constant $|$SPREAD\_MODEL$|$=0.003, while the red dotted lines are
  delineating the region used to select stars
  $|$SPREAD\_MODEL$|<$0.003+SPREADERR\_MODEL}
\label{fig:stargal}
\end{figure}
\begin{figure}
\includegraphics[]{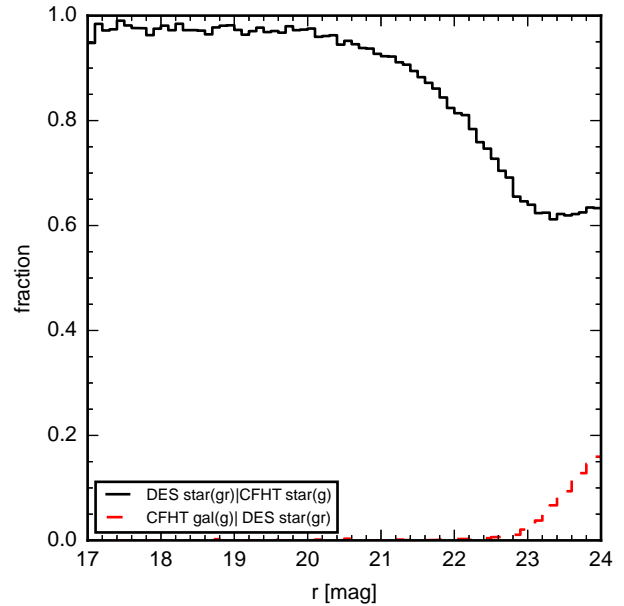}
\caption{Quality of the star-galaxy separation as inferred using the
  SPREAD\_MODEL parameter. The photometry extracted from the DES
  images is compared to the star-galaxy separation in the CFHT Wide
  survey of the same area on the sky. Black solid curves show the
  fraction of objects classified as likely stars by CFHT CLASS\_STAR
  $>$0.5 that are also classified as stars according to the
  SPREAD\_MODEL cut (Eq.~\ref{eq:sg}). This is a measure of
  completeness of our stellar sample. Red dashed curve shows the
  fraction of objects classified as stars according to the
  SPREAD\_MODEL cut but are classified as likely galaxies by CFHT
  CLASS\_STAR $<$ 0.5. This is a measure of the contamination.}
\label{fig:stargal_check}
\end{figure}

\section{DES Data Processing}

\begin{figure*}
\includegraphics[width=0.98\textwidth]{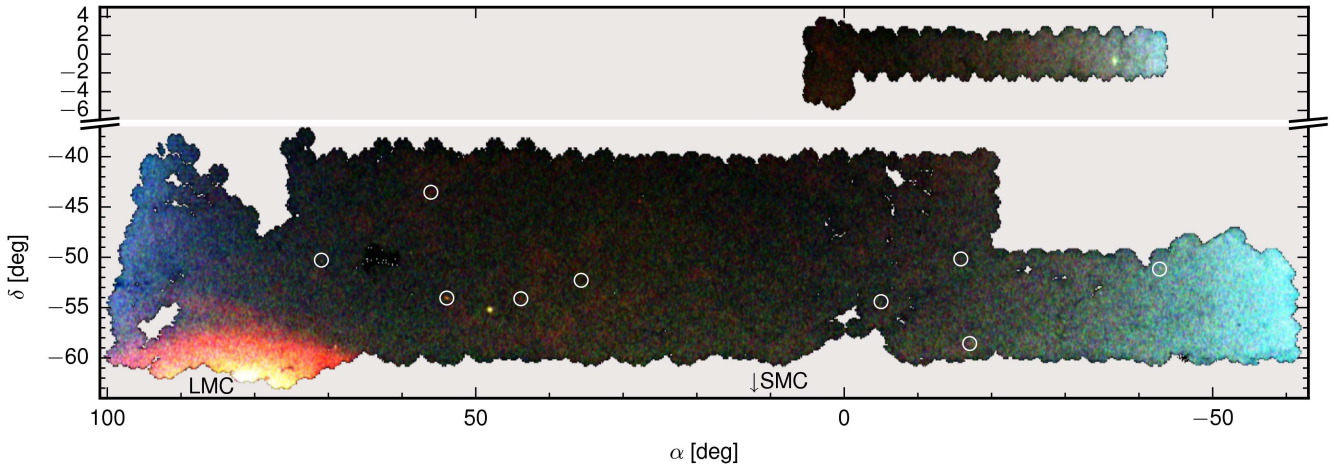}
\caption{False-colour map of the density of the main sequence turn-off
  (MSTO) stars with 0.2$<$g$-$r$<$0.6 as observed by the DES
  survey. The density in different magnitude bins corresponding to
  different distances is used to create different colour channels:
  blue 17$<$r$<$19 (corresponding to distances of $\sim$ 4$-$10\,kpc),
  green 19$<$r$<$21 (10$-$25\,kpc), red 21$<$r$<$22.75
  (25$-$56\,kpc). The MW satellites discovered in this paper are shown
  with white unfilled circles. Note that the declination axis has been
  broken to avoid showing large empty area between
  -35$\lesssim\delta\lesssim$-5 without DES data}
\label{fig:colormap}
\end{figure*}

DES enjoys the use of the 2.2-degree field-of-view of the 570
megapixel DECam camera mounted on the 4m Blanco telescope at Cerro
Tololo in Chile to obtain deep images of an extensive swath of the
Southern sky around the Magellanic Clouds. DES is an international
collaboration, whose primary focus is on the extra-galactic
science. DES started taking data in August 2013 and will continue for
five years. Currently, all its data products are proprietary, but the
individual processed images are eventually released to the community
through the NOAO infrastructure with a 1 year delay. In this paper, we
present the results of the analysis of the entire DES imaging data
released to date, i.e. $\sim$2100 square degrees obtained in the
$1^{\rm st}$ year of operations. Here, we concentrate on what clearly
is a by-product of the DES survey: the stellar photometry and
astrometry.

The NOAO Science
Archive\footnote{\url{http://www.portal-nvo.noao.edu/}} hosts both raw
and processed DES images.  We have queried the database through the
interface provided and selected all available InstCal calibrated
images, the corresponding weight maps and the masks\footnote{The exact
  query used to select images from the NOAO Science archive: SELECT
  reference, dtpropid, surveyid, release\_date, start\_date,
  date\_obs, dtpi, ra, dec, telescope, instrument, filter, exposure,
  obstype, obsmode, proctype, prodtype, seeing, depth, dtacqnam,
  reference AS archive\_file, filesize, md5sum FROM voi.siap WHERE
  release\_date $<$ '2015-01-18' AND (proctype = 'InstCal') AND
  (prodtype IS NULL OR prodtype $<>$ 'png')) AND (dtpropid ILIKE
  '\%2012B-0001\%' OR surveyid ILIKE '\%2012B-0001\%') }. The images
were first staged for the download, and subsequently copied across for
further processing. The total size of the released images, weight maps
and masks in Rice compressed format was $\sim$4.6 TB.  Once
uncompressed, the size of the images and weight maps increases by a
factor of $\sim$ 10. For the analysis reported below, only $g$,$r$ and
$i$ band images were analysed ($\sim$ 30\,TB of data).

DES images are taken with 90s exposure and the DECam pixel
scale is 0.26$^{\prime \prime}$/pixel. InstCal products, delivered by
the DECam pipeline, are single-frame images (not stacks) that have
been bias, dark and flat-field calibrated as well as cross-talk
corrected. Additionally, defects and cosmic rays have been masked and
the WCS astrometry provided\footnote{Detailed information available
  through the DECam Data Handbook at \url{http://bit.ly/1D7ZGDc}}.

\subsection{From Images to Catalogues}

For the subsequent catalogue creation, we have relied heavily on the
SExtractor/PSFEx routines \citep[see][for more
  details]{sextractor,psfex,annu13}. Typically, processing a single
frame consisting of 60 CCD chips with SExtractor/PSFEx took $\sim$ 2.5
CPU-hours, leading to a total budget for the three DES filters of
$\sim$15000 CPU-hours. The image processing has been carried out using
the Darwin HPC computing facility\footnote{
\url{http://www.hpc.cam.ac.uk}} as well as a local 8 node/96 core
computer cluster. The main catalogue assembly steps are as follows:

\begin{itemize}

\item Initial SExtractor pass. To provide the starting parameters for
  the PSFEx routine, the DES images are analysed by SExtractor. The
  flag settings are standard, albeit the \verb!PHOT_APERTURES! flag is
  adjusted to match the value of the \verb!G_SEEING! keyword in the
  header (if it is available). Additionally, the downloaded weight
  maps are provided to SExtractor via the \verb!WEIGHT_MAP! option.

\item PSFEx run. The output of SExtractor is used to calculate the
  PSFs for the subsequent photometry.  PSFEx is run on each CCD chip
  separately with default PSFEx options, except the \verb!PSF_SIZE!
  keyword which is set to (40,40) pixels.

\item Final SExtractor pass. Using the PSF models computed in the
  previous step, SExtractor is run again to determine {\it Model} and
  {\it PSF} magnitudes.  This step is the longest in the entire
  sequence.

\item Ingestion. For each filter, the resulting SExtractor catalogues
  are ingested into a PostgreSQL database as separate tables. \verb!Q3C!
  spatial indices \citep{koposov06} are created to speed up further
  steps.

\item Duplicate removal. Due to significant image overlaps, sources
  appearing more than once have to be flagged in the catalogues. For
  each source we perform a search with 1$\arcsec$ radius, and mark all
  objects coming from different frames and HDUs as secondary.

\item Photometric calibration. To convert instrumental magnitudes into
  calibrated ones for each frame (including multiple HDUs) a
  cross-match between the DES and the APASS DR7 survey data is
  performed. The zero-point is then measured as a median offset with
  respect to the APASS photometry on a per-field basis. The resulting
  photometric precision as estimated using overlaps of different
  frames is: 0.03 mag (Gaussian $\sigma$) in the g-band, 
 0.025 mag in the r-band, and 0.033 mag in the i-band.

\item Band-merging. The lists of primary sources in the $g$, $r$ and
  $i$ bands are combined into the final catalogue using the matching
  radius of 1 arcseconds. For the purpose of this paper, the catalogue
  was based on $r$-band, i.e. $r$-band detection was required in order
  for the object to be in the final catalogue.

\item The final \verb!Q3C! index is created on the band-merged catalogue to 
  speed up spatial searches in the dataset.

\end{itemize}

\subsection{Photometry and star-galaxy separation}

\input{table.tex}

PSF magnitudes of star-like objects are given by the \verb!MAG_PSF!
  output of SExtractor. As an indicator of star-galaxy separation we
  use the \verb!SPREAD_MODEL! parameter provided by SExtractor. This
  is a metric similar to \verb!psfmag-modelmag! used by SDSS (see
  Fig.~\ref{fig:stargal}). A sensible selection threshold for bright
  stars would be $|\verb!SPREAD_MODEL!|<0.003$ \citep{desai12,annu13}, however
  for faint magnitudes this cut causes significant incompleteness in
  stars. Therefore, instead we choose to
  require:\footnote{\url{http://1.usa.gov/1zHCdrq}}
\begin{equation}
|\verb!SPREAD_MODEL!|<0.003+\verb!SPREADERR_MODEL!\label{eq:sg}
\end{equation}
This particular cut ensures that the stellar completeness remains
reasonably high at faint magnitudes, while the contamination is kept
low at the same time. The behaviour of $0.003+\verb!SPREADERR_MODEL!$
as a function of magnitude shown in Figure~\ref{fig:stargal} explains
why a fixed \verb!SPREADERR_MODEL!  threshold is suboptimal. To assess
the levels of completeness and contamination induced by our stellar
selection, we use a portion of the DES-covered area of sky overlapping
with the CFHTLS Wide survey \citep{hudelot12}. This is a dataset of
comparable depth, for which morphological object classifications are
provided. Figure~\ref{fig:stargal_check} gives the resulting
performance of the stellar selection procedure in which Equation
~\ref{eq:sg} is applied to both $g$ and $r$-band catalogues.  In
particular, the Figure gives completeness (black solid histogram)
calculated as the fraction of objects classified as stars by CFHTLS
(their \verb!CLASS_STAR!$>$0.5) which are also classified as stars by
our cuts applied to the DES data. Similarly, contamination can be
gleaned from the fraction of objects classified as galaxies by the
CFHTLS but as stars by our DES cuts (red dashed line). It is
reassuring to observe low levels of contamination all the way to the
very magnitude limit of the DES survey. At the same time, completeness
is high across a wide range of magnitudes and only drops to $\sim$
60\% for objects fainter than $r\sim22$. It is also worth noting that
the star-galaxy separation criteria employed in this work may not be
ideally suited for other studies, as they may have different
requirements in terms of the balance between the completeness and the
contamination.

In the stellar catalogues built using the procedure described above,
the magnitudes are equivalent to the SDSS $gri$. Consequently, the
extinction coefficients used are those suitable for the SDSS
photometric system, while the dust reddening maps employed are from
\citet{schlegel98}. Note that the depth of the resulting catalogues
varies somewhat across the DES footprint, but could be approximately
estimated from the source number counts in $g$, $r$, $i$
filters. These number counts peak at magnitudes 23.7, 23.6, 22.9 in
$g$, $r$, $i$ correspondingly, indicating that the catalogues start to
be significantly affected by incompleteness at somewhat brigher magnitudes 
g$\sim$23.5, r$\sim$23.4, i$\sim$22.7.

To illustrate the quality of the resulting catalogue,
Figure~\ref{fig:colormap} displays the density of the Main Sequence
Turn-Off (MSTO; 0.2$<$g$-$r$<$0.6) stars on the sky.  The density of
stars with 19$<$r$<$21 (corresponding to distances of 10$-$25\,kpc) is
shown in the green channel, more distant stars with 21$<$r$<$22.75
(corresponding to distances of 25$-$56\,kpc) are used for the red
channel, and the nearby stars with 17$<$r$<$19 (distances of
4$-$10\,kpc) in the blue channel. This map is an analog of the "Field
of Streams" picture by \citet{belokurov06}. The density distribution
is very uniform thus confirming the high precision and the stability
of the photometry as well as the robustness of the star-galaxy
separation across the survey area. The map also reveals some of the
most obvious overdensities discovered in this work, at least two of
which are visible as bright pixels in the Figure.

\begin{figure*}
\begin{center}
\includegraphics[width=0.95\textwidth]{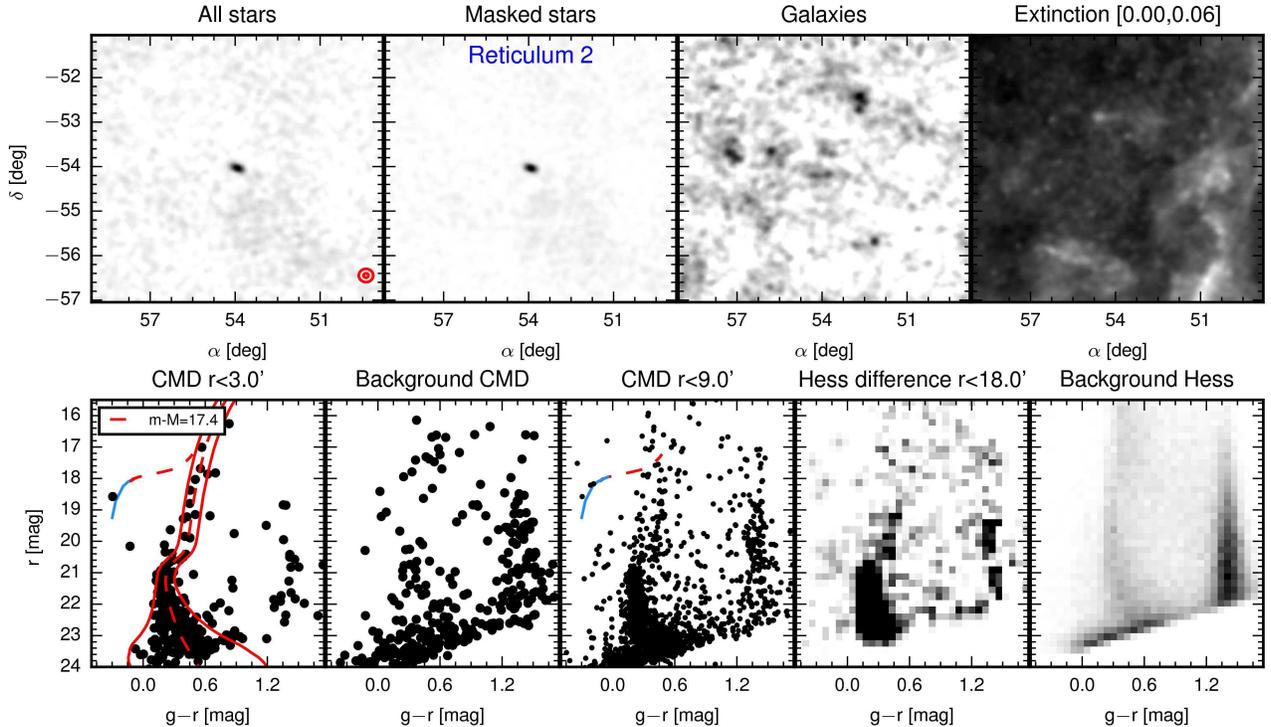}
\end{center}
\caption{Reticulum 2 satellite.{\it Top row, 1$^{\it st}$ panel:} Density of
  all stars centered on the satellite's location, dark shades of grey
  correspond to density enhancement. Red circles show the size of the apertures
  used to create CMDs given in the 1st and the 3rd bottom panels. Note
  the prominent over-density in the centre.{\it Top row, $2^{\it nd}$ panel:}
  Density of stars selected using the isochrone mask shown in the 1st
  bottom panel. The background/foreground density variations have now
  subsided further as compared to the left-most panel. {\it Top row,
    3$^{\it rd}$ panel:} Density of galaxies in the vicinity of the
  satellite. No obvious over-density coincident with the object is
  visible. {\it Top row, 4$^{\it th}$ panel:} Distribution of Galactic dust
  reddening $E(B-V)$ according to the measurement of
  \citet{schlegel98}. The extinction values corresponding to black and white colors in the panel are given in the panel title. {\it Bottom row, 1$^{\it st}$ panel:} Colour-Magnitude
  distribution of stars inside the smaller of the two apertures shown
  in the 1st Top panel, centered on the object. Note the unmistakable
  overdensity corresponding to the Main Sequence and Main Sequence
  Turn-Off stars. The Red Giant Branch stars are also visible, albeit
  more scarce. There appears to be at least one possible Blue
  Horizontal Branch star member, as indicated by the good match
  between with isochrone (red dashed line) and the M92 BHB ridge-line
  (blue solid line). {\it Bottom row, 2$^{\it nd}$ panel:} Comparison CMD built
  from random nearby stars outside the larger aperture, but covering
  the same on-sky area as in leftmost bottom panel. The comparison
  shows various stellar pile-ups, most prominently the disk dwarfs and
  objects near the detection limits, but features similar to that of
  the satellite's populations are not observed. {\it Bottom row,
    3$^{\it rd}$
    panel:} Satellite's CMD for stars within the larger of the two
  apertures shown in the 1st top panel. At least 3 likely BHB members
  are visible. {\it Bottom row, 4$^{\it th}$ panel:} Hess difference for stars
  within and outside the aperture of $r=18^{\prime}$. Note the
  satellite's MS, MSTO and RGB populations standing out clearly. {\it
    Bottom row, 5$^{\it th}$ panel:} Density of nearby background stars in the
  colour-magnitude space (Hess diagram).}
\label{fig:reticulum2}
\end{figure*}

\section{Search for stellar over-densities}

To uncover the locations of possible satellites lurking in the DES
data, we follow the approach described in
\citet{koposov08,walsh09}. In short, the satellite detection relies on
applying a matched filter to the on-sky distributions of stars
selected to correspond to a single stellar population at a chosen
distance. The matched filter is simply a difference of 2D Gaussians,
the broader one estimating the local background density, while the
narrow one yielding the amplitude of the density peak at the location
of the satellite.

We start by taking a catalogue of sources classified as stars. A
sub-set of these is then carved out with either a set of
colour-magnitude cuts or with an isochrone mask offset to a trial
distance modulus.  Then a 2D on-sky density map of the selected stars
is constructed, keeping the spatial pixel sufficiently small,
e.g. 1$^{\prime}$ on a side. At the next step, the density map is
convolved with a set of matched filters (described above) with
different inner and outer kernels. Finally, these convolved maps are
converted into Gaussian significance maps and the most significant
over-densities are extracted.

In the analysis presented here, we have used the mask based on the
PARSEC isochrone \citep{bressan12} with an age of 12 Gyr and
metallicity of
${\rm [Fe/H]}=-2$ \footnote{\url{http://stev.oapd.inaf.it/cmd}}, spread out by
the characteristic DES photometric error as a function of colour and
magnitude. The trial distance moduli explored in the search range from
m$-$M=15 to m$-$M=24. The inner kernel size is allowed to vary from
1$\arcmin$ to 10$\arcmin$. After extracting the most significant
over-densities all detections are cross-matched with the list of
positions of known dwarf galaxies from \citet{mcconnachie12}, nearby
LEDA galaxies \citep{paturel03} with radial velocities less than
2000\,km/s, globular clusters in the catalogue of \citet{harris96} as
well as the globular clusters listed in Simbad.

\begin{figure*}
\begin{center}
\includegraphics[width=0.95\textwidth]{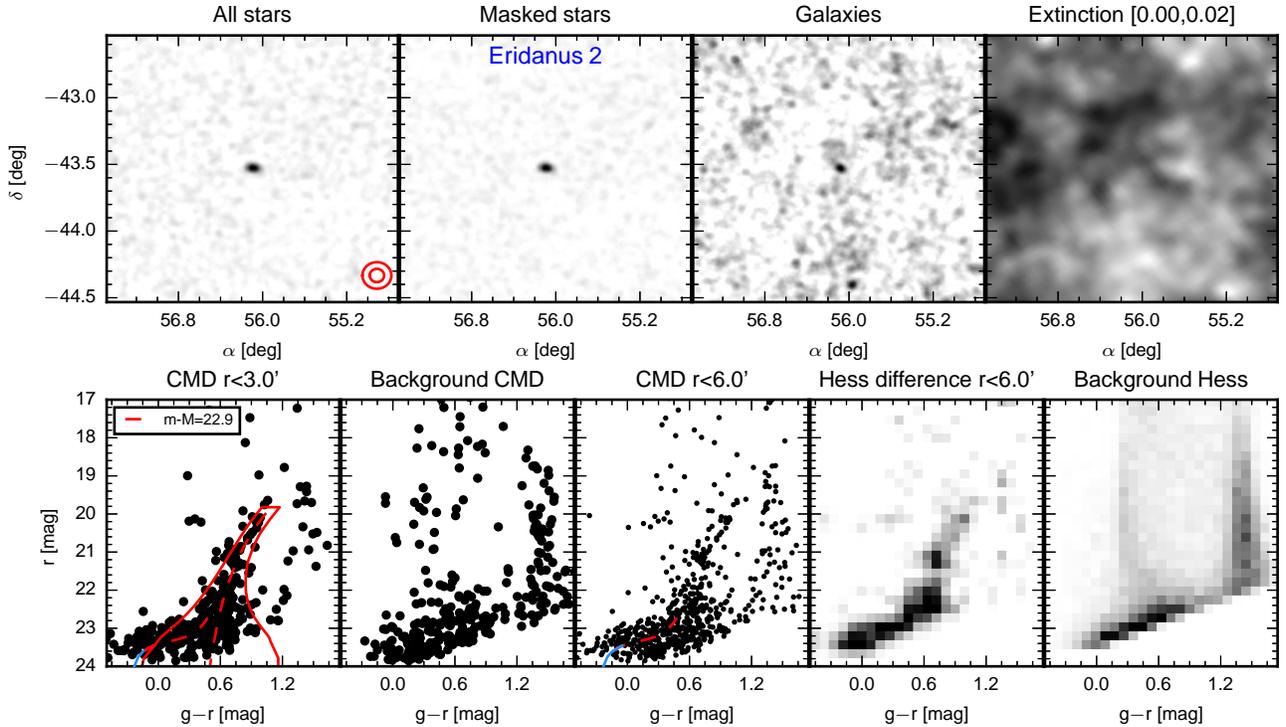}
\end{center}
\caption{Eridanus 2. See Figure~\ref{fig:reticulum2} for detailed
  description of the panels. This is the most distant of the detected
  satellite. Note the faint over-density of objects classified as
  galaxies at the location of the satellite. These are mis-classified
  blended stars unresolved by DES+SExtractor. The CMD shows prominent
  RGB and BHB with a handful of possible young Blue Loop stars.}
\label{fig:eridanus2}
\end{figure*}

The list of over-densities produced is then cleaned for duplicates,
e.g. over-densities detected with more than one distance modulus
and/or kernel size. We have also removed detections near the very edge
of the survey's footprint. The remaining candidates are ranked by
their significance and eye-balled. Among the 9 objects presented in
this paper, 5 are at the very top of the ranked list with
significances well above 10$\sigma$, while a further 4 are somewhat
lower in the list (but with significances of order of 10$\sigma$). We
firmly believe that all objects presented below are genuine new
satellites of the Milky Way. Apart from the over-density significance,
our inference is based on a combination of additional factors,
including their morphological and stellar population properties. We
describe this overwhelming observational evidence in detail in the
next Section. The list of objects and their properties is given in
Table~\ref{tab:params} and discussed in detail in the next Section.

\section{Properties of the detected objects}

While our search for stellar over-densities returned a large number of
promising candidates, in this Paper we concentrate on a particular
{\it gold-plated} sample of objects whose nature can be unambiguously
established with the DES data alone. Let us briefly review the sanity
checks carried out to corroborate the classification. First, each of
the nine objects possesses a significant stellar overdensity, in fact,
the significance is in excess of 10$\sigma$ in eight out of nine
cases. Second, no obvious background galaxy overdensity is visible,
thus reducing dramatically the possibility that the objects are
spurious density peaks caused by mis-classified faint galaxies. Third,
the dust maps in the vicinity of each of the satellite do not show any
strong features that can be linked to the stellar overdensity. Most
importantly, distributions of stars in the colour-magnitude plane
reveal sequences corresponding to coeval populations at the same
distance. In all cases, not just one particular stellar population is
visible but several, for example, MSTO+RGB+BHB.

\begin{figure*}
\begin{center}
\includegraphics[width=0.95\textwidth]{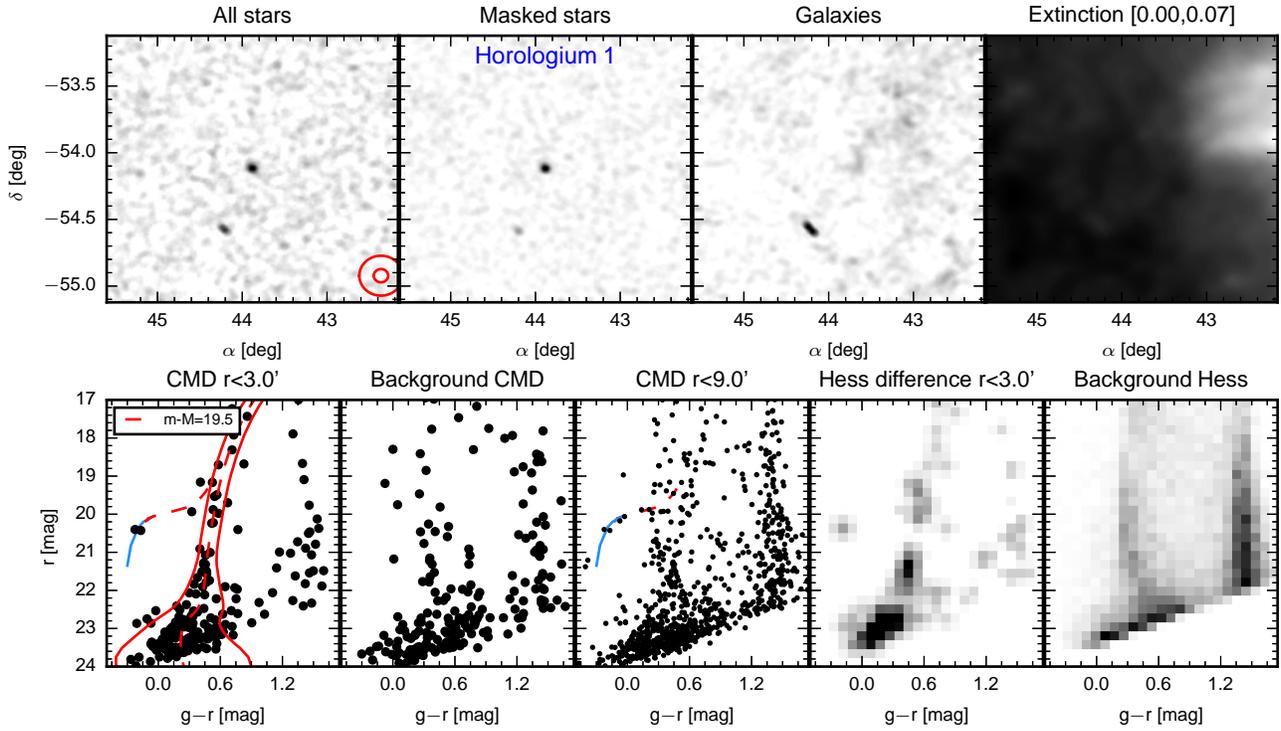}
\end{center}
\caption{Horologium 1 satellite. See Figure~\ref{fig:reticulum2} for
  detailed description of the panels. The CMD reveals an obvious MSTO
  and RGB. There are also several likely BHB members.}
\label{fig:horologium1}
\end{figure*}

\begin{figure*}
\begin{center}
\includegraphics[width=0.95\textwidth]{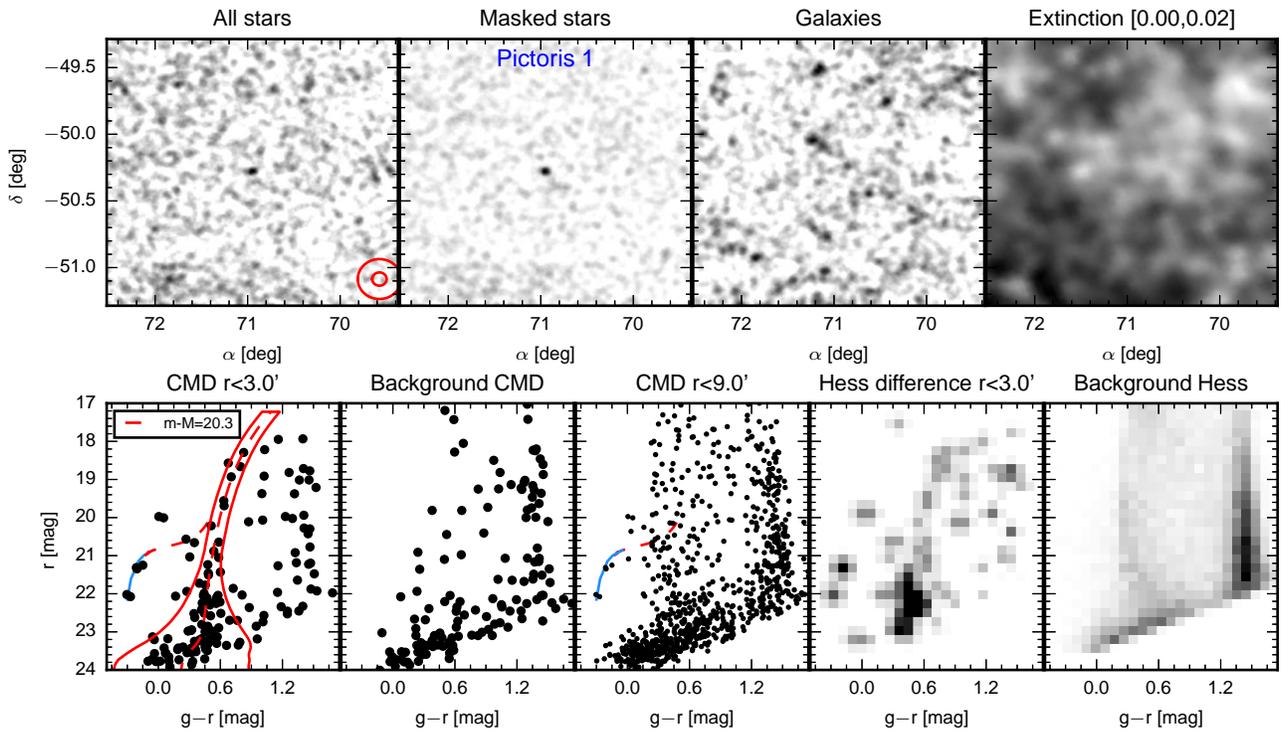}
\end{center}
\caption{Pictoris 1 satellite. See Figure~\ref{fig:reticulum2} for
  detailed description of the panels. Pictoris 1 resembles Horologium
  1 (see Figure~\ref{fig:horologium1} albeit located at a slightly
  larger distance. Therefore the MSTO is barely detected as it kicks
  in right around the DES limiting magnitude.}
\label{fig:pictoris1}
\end{figure*}

\begin{figure*}
\begin{center}
\includegraphics[width=0.95\textwidth]{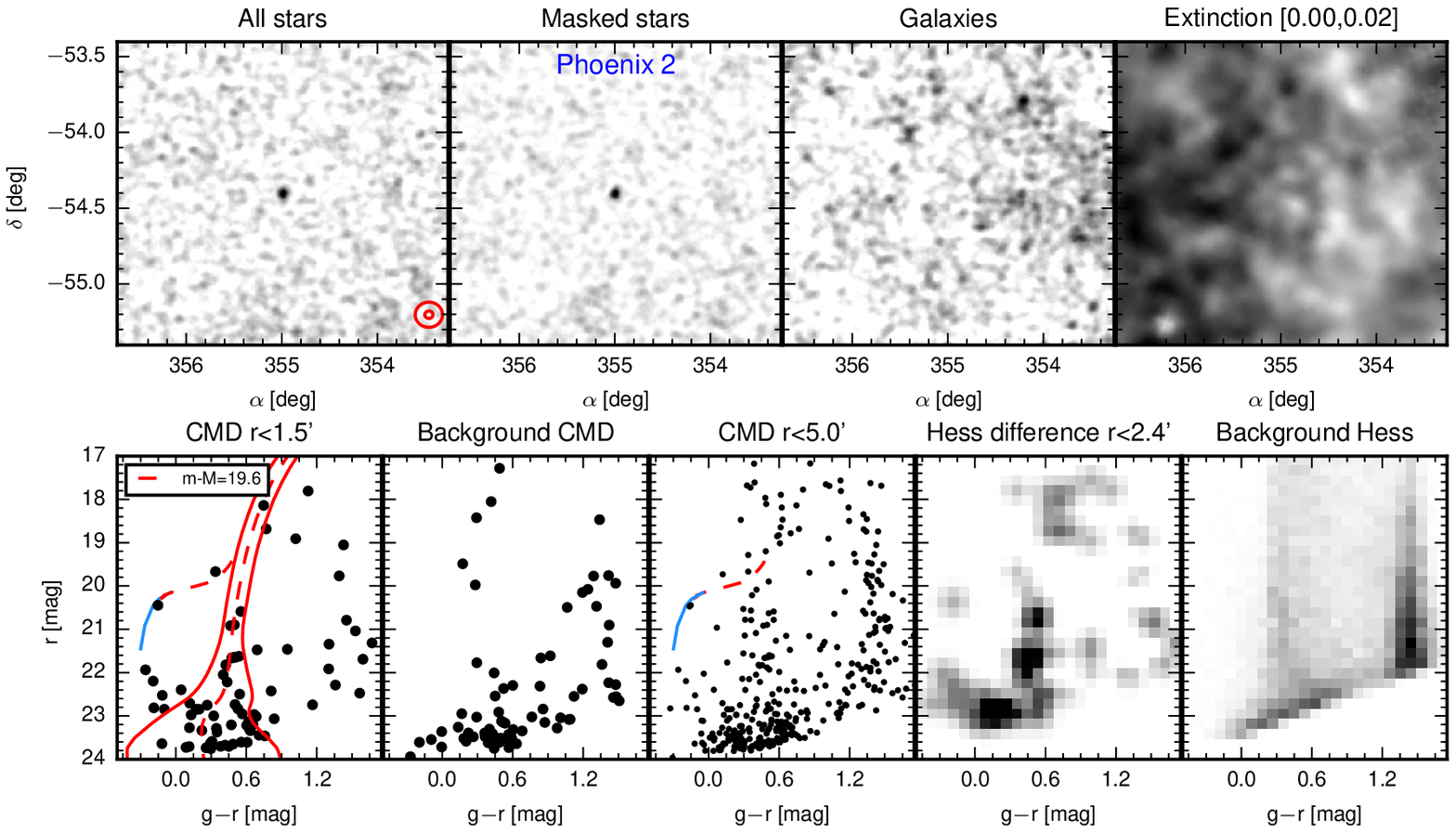}
\end{center}
\caption{Phoenix 2 satellite. See Figure~\ref{fig:reticulum2} for
  detailed description of the panels. The satellite looks similar to
  Horologium 1 and Pictoris 1, but with a noticeable Blue Straggler
  sequence.}
\label{fig:phoenix2}
\end{figure*}

Figures~\ref{fig:reticulum2}-\ref{fig:tucana2} give the visual summary
of the properties of each of the detected satellites and serve to
substantiate our classification. Each of these diagnostic plots
comprises of 9 panels: 4 in the top row are concerned with the on-sky
distribution of the object's stars, while the 5 in the bottom row give
details of its stellar populations. In particular, the density of all
objects classified as stars is shown in the top left panel. The second
panel in the row gives the density of all stars selected with the
best-fit isochrone mask (shown in the left bottom panel). For
comparison, the density of objects classified as galaxies is presented
in the third panel, and the galactic extinction map from
\citet{schlegel98} in the fourth. 

The first panel in the bottom row displays the colour-magnitude
distribution of stars selected to lie in the small aperture (whose
size is indicated by the small red circle in the bottom right corner
of the top left panel) centered on the object. Red solid lines are the
boundaries of the isochrone mask used to select the probable satellite
member stars, while the isochrone itself is shown as a red dashed
line. The horizontal branch (HB) in the model isochrone does not
extend to sufficiently blue $g-r$ colours, therefore we complement the
theoretical HB with a fiducial HB (blue solid line) built using data
for the metal-poor old globular cluster M92 from
\citet{bernard14}. The object's CMD can be contrasted to the
foreground/background CMD shown in the second bottom panel. This is
built with random nearby stars located within large annulus 
(typical inner radius 0.2$\degr$ and outer radius of 1$\degr$) and sampling the same number of
stars as that in the left panel. The aperture chosen to create the CMD
in the left panel is sufficiently small in order to minimize the
contamination by non-member stars. This, however, implies that some of
the rarer stellar populations (such as BHBs for example) are not fully
sampled. A more complete view (albeit with a higher contamination) of
the satellite's stellar populations can be seen in the third bottom
panel for which a larger aperture is used (large red circle in the top
left panel). In most cases, this particular view of the CMD
strengthens the detection of the BHB population. The last two panels
in the bottom row are Hess diagrams, i.e. density distributions in the
colour-magnitude space. In particular, the fourth panel is the Hess
difference between the stars inside a small aperture centered on the
satellite (the aperture size is shown in the panel's title) and a
large area well outside the satellite's extent. Finally, for
comparison, the Hess diagram of the background/foreground population
is shown the fifth panel of the bottom row.

All objects we report in Table~\ref{tab:params} have passed the sanity
checks described above with flying colours. Moreover, some of the new
satellites are actually visible in the DES image cutouts as described
in the sub-section~\ref{sec:notes}.

{ The naming convention we have adopted uses the constellation of
  the object together with the corresponding arabic numeral. This
  differs from the customary bimodal naming convention for globular
  clusters and dwarf galaxies, that leads to the name dependance on
  the knowledge of whether the object is a cluster or a galaxy.}

\subsection{Satellite modelling}

Having identified the objects of interest, we proceed to measuring
their distances, luminosities and structural parameters. First, the
distances are determined using the available BHB stars if present. We
have used both the theoretical HB isochrone corresponding to the old
metal-poor population mask used for the selection of the likely
satellite members and the M92 HB fiducial from \citet{bernard14}
converted into the SDSS photometric system using the equations from
\citet{tonry12}. The distance moduli implied by these HBs are
consistent within 0.1 mag. { To verify the distance accuracy 
we also fitted the Hess diagram of each object with a stellar population model 
\citep[see e.g.][for more details]{koposov10}. For all objects (excluding Tucana~2) this gave us 
similar distance modulus differing by less than 0.1 - 0.2 mag. For Tucana~2 we use the distance implied by the Hess diagram fit.}


%

To measure the morphological properties of all the detected objects, we
have modelled the on-sky 2D distribution of the likely member stars
\citep[see e.g.][for similar approach]{martin08}. For each object,
only stars located within the CMD mask are used for the morphological
analysis. The model used to describe the density distribution of stars
is a mixture of the background/foreground stellar density (assumed to
be constant across a few degrees around the object) and a rotated
elongated exponential model for the satellite's stars. Therefore, the
probability of observing a star at a particular position on the sky
with coordinates $\alpha$ and $\delta$ is:

$$ P(\alpha, \delta|\theta) = ( 1 - f)\, P_{bg} (\alpha, \delta|\theta) + f
P_{obj}(\alpha,\delta|\theta)$$ 
$$ P_{bg}(\alpha, \delta|\theta) = \frac{1}{\Omega\,\cos \delta_c}$$
$$ P_{obj}(\alpha, \delta|\theta) = \frac{1}{2\pi\,(1-e)\,\cos \delta_c\,a_{maj}^2} \exp \left({-\frac{D_{ell}}{a_{maj}}}\right) $$

and $$D_{ell} = \left| \left| \begin{pmatrix} \frac{\cos (PA)}{(1-e)} &
  \frac{\sin(PA)}{(1-e)}\\ -\sin(PA) &
  \cos(PA) \end{pmatrix} \begin{pmatrix}(\alpha-\alpha_c)\,\cos \delta_c\\ \delta-\delta_c\end{pmatrix}
  \right|\right|$$

Here $\Omega$ is the area of the analysed part of the data and $\theta$ is
a shorthand notation for all model parameters, of which there are 6 in
total. These are: $f$ - the fraction of objects belonging to the object,
$a_{maj}$ -- the major axis (the exponential scale length), $e$ --
ellipticity of the object, $PA$ positional angle of the major axis,
and finally, $\alpha_c\,\delta_c$ -- the position of the center of the
object.

The above model is fit to the data assuming uninformative priors on
$\alpha_c,\delta_c$, flat priors on $e$ and PA, (i.e. e $\sim$
Uniform(0, 0.8), PA $\sim$ Uniform(0, 180)) and the Jeffreys priors on
$f$ and $a_{maj}$, namely $f \sim$ Beta(0.5, 0.5) and $P(a_{maj})
\propto \frac{1}{a_{maj}}$ \citep[see e.g.][]{mackay02,gelman14}. The
resulting posterior is sampled using the Affine invariant Ensemble
sampler \citep{goodman10,foreman13}. The values of parameters and the
associated uncertainties are given in Table~\ref{tab:params}.  To
check whether the ellipticity of an object is significant, the Bayes
Factor is calculated using the Savage-Dickey ratio
\citep{verdinelli95,trotta07}.  As the Table~\ref{tab:params} reports
(final column), only Reticulum~2 and Eridanus~2 show strong,
statistically significant elongation. Other objects are consistent
with being circular.

Having modelled the morphology of each satellite, the posterior
distributions of their structural paramereters are used to estimate
the total number of stars belonging to the satellite. The satellite
luminosities are calculated only with stars with r$<$23 falling within
the isochrone mask (we have used a brighter magnitude limit to avoid
incompleteness due to the star/galaxy misclassification). Then, the
number of stars inside the isochrone mask is converted into the
stellar mass and the corresponding total luminosity assuming an old
metal-poor isochrone and the KTG/Chabrier IMF \citep{ktg,chabrier03}.

Figure~\ref{fig:profiles} is a gallery of 1D azimuthally averaged
stellar density profiles for each of the satellites. Red lines are the
best fit exponential models. Before we discuss the size-luminosity
distribution of our sample in sub-section~\ref{sec:mv_rh}, let us have
a closer look at the individual properties of each of the satellites.

\subsection{Notes on individual objects}
\label{sec:notes}

\begin{figure*}
\begin{center}
\includegraphics[width=0.95\textwidth]{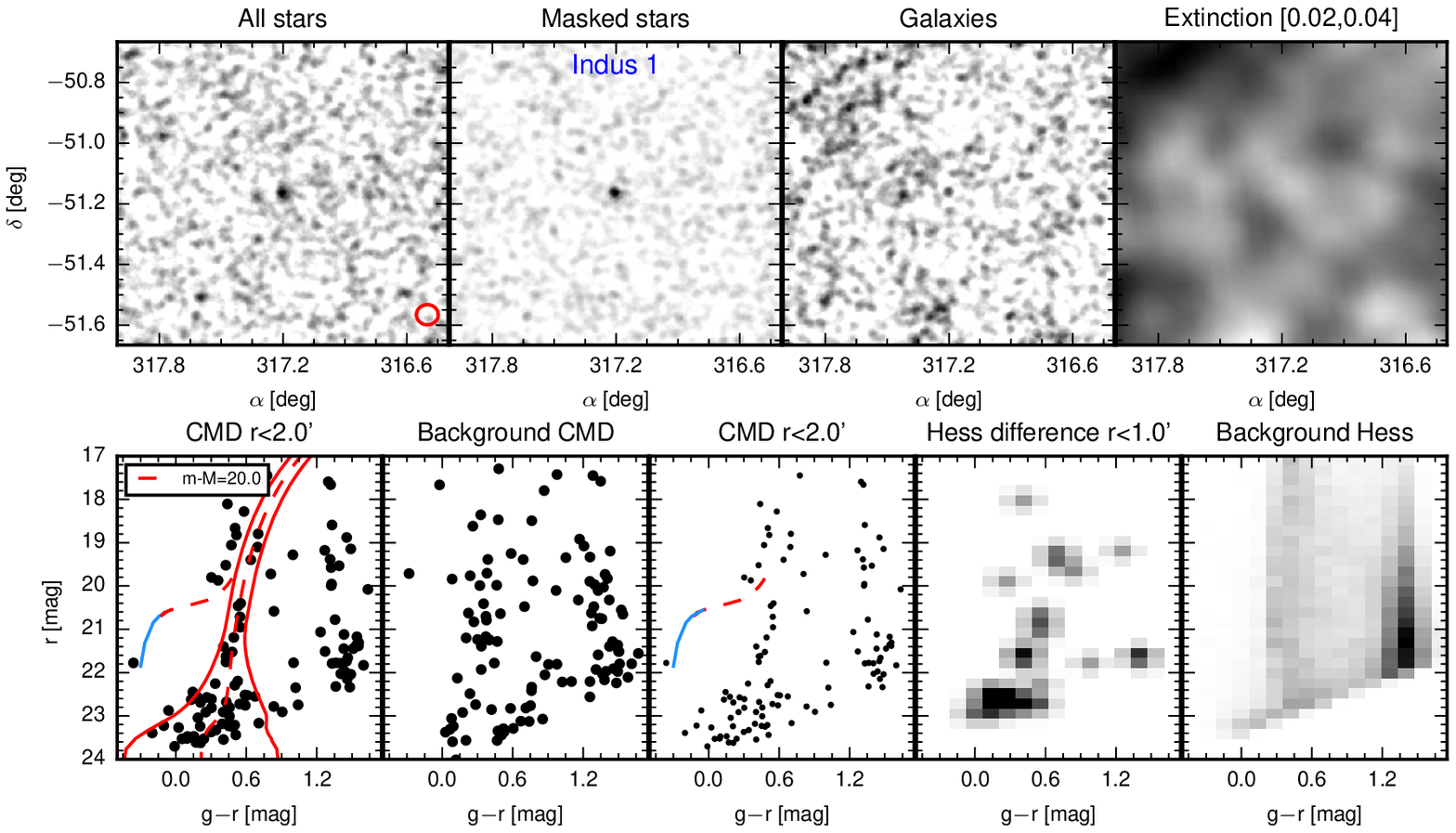}
\end{center}
\caption{Indus 1 satellite. See Figure~\ref{fig:reticulum2} for
  detailed description of the panels. Located at a similar distance to
  Horologium 1, Pictoris 1 and Phoenix 2, Indus 1 is more compact and
  boasts only one possible BHB star. Curiously the satellite is
  visible in the DES image-cutout presented in the companion
  Figure~\ref{fig:indus1_eridanus3_color}.}
\label{fig:indus1}
\end{figure*}

\begin{itemize}

\item Reticulum~2 (see Figure~\ref{fig:reticulum2}) -- With the
  significance in excess of 40, this is the most obvious previously
  unknown object in the DES field of view. The CMD shows clear MS and
  MSTO together with beginnings of a RGB. Reticulum 2 is clearly very
  elongated: the measured axis ratio is around 0.6. It is one of the
  two objects in the sample presented here, for which the flattening
  of the stellar density is statistically significant. Its luminosity
  is $M_V=-2.7$ and the half-light radius of 30 pc, which brings it
  just outside the cloud of classical globular clusters into either
  extented globular clusters or dwarf galaxies. It is impossible to
  classify this satellite securely with the data in hand, therefore we
  shall designate this and similar objects as "ultra-faint
  satellites". Clearly, kinematic and chemical analysis is required
  before the nature of this object can be established.  It is
  reasonable to associate the low axis ratio measured for this object
  with the effects of the Galactic tides. However, at the moment, we
  do not have any firm evidence of extra-tidal material outside the
  satellite. The distance estimate for Reticulum 2 is 30\, kpc as
  judged by a couple of BHB stars as well as the prominent MS and
  RGB. Note that eight other satellites are all substantially further
  away.
 
\item Eridanus~2 (see Figure~\ref{fig:eridanus2}). This is the second
  most significant and the most distant object in our list. With its
  location on the periphery of the Galaxy at $\sim$ 380\,kpc,
  luminosity of $M_V=-6.6$ and the size of $\sim$ 170\,pc, this appears
  to be a twin of the Leo~T dwarf galaxy \citep{irwin07}. Similarly to
  Leo~T, Eridanus~2 is clearly visible in the DES color images
  (Figure~\ref{fig:eridanus2_color}) as a group of faint blue stars
  embedded in a blue low surface-brightness cloud. In contrast to
  Leo~T though, Eridanus~2 shows significant elongation with a
  measured ellipticity of $\sim$ 0.4. Its color-magnitude diagram
  boasts a strong RGB and a very prominent red clump. Additionally,
  there appears to be a group of bright blue stars which could
  possibly be interpreted as young blue-loop stars (see
  Figure~\ref{fig:iso_eridanus2}), similar to those found in
  Leo~T \citep{weisz12}. The HB stars of Eridanus~2 are just at the
  magnitude limit achievable with the single-frame DES photometry. The
  possible presence of young stars and large distance to the
  Eridanus~2 galaxy suggest that the satellite is a viable candidate
  for follow-up HI observations. The color images of the galaxy (see
  both panels of Figure~\ref{fig:eridanus2_color} also show an curious
  fuzzy object which can be interpreted as a very faint globular
  cluster. The object is almost at the center of the galaxy
  ($\alpha,\delta$)=($56.0930,-43.5327$) and is very diffuse with a size
  consistent with a few parsecs assuming it is at distance of
  Eridanus~2\footnote{There is another compact but high surface
    brightness round object looking consistent with a globular cluster
    ($\alpha,\delta$)= (56.0475,-53.5213) but it is probably a
    background galaxy as even its outskirts do not resolve into
    stars}. If any of the faint diffuse objects superposed with
  Eridanus~2 are indeed globular clusters, this would make this
  satellite the faintest galaxy to possess stellar sub-systems!

\begin{figure*}[t]
\begin{center}
\includegraphics[width=0.95\textwidth]{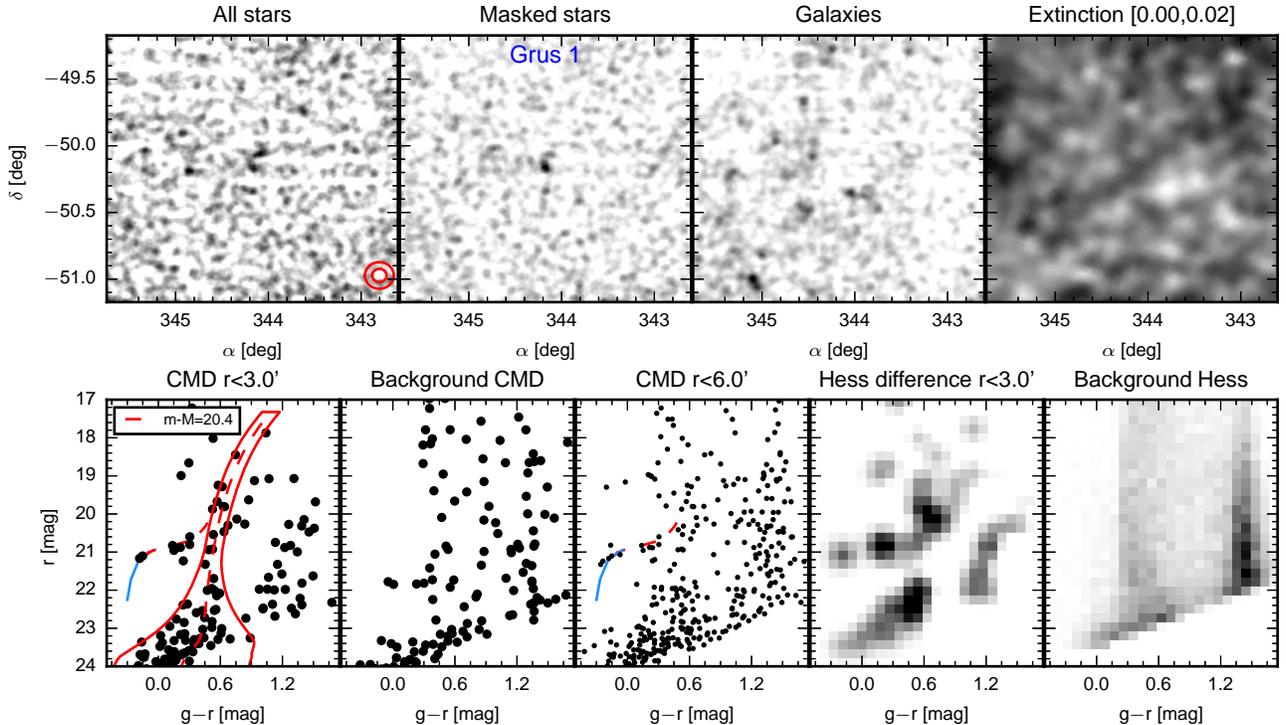}
\end{center}
\caption{Grus 1. See Figure~\ref{fig:reticulum2} for detailed
  description of the panels. Grus 1 is situated at a distance similar
  to that of Horologium 1, Pictoris 1, Phoenix 2 and Indus 1. The
  central stellar overdensity is not as prominent as in the previously
  discussed objects. This is partly explained by the fact that Grus 1
  falls right into the CCD chip gap. Nonetheless, the CMD shows off
  such familiar features as an RGB and a well-populated HB.}
\label{fig:grus1}
\end{figure*}

\item Horologium~1 (see Figure~\ref{fig:horologium1}). This is the
  third most obvious object within the DES field-of-view with a
  significance of $\sim$ 30.  Horologium~1 is one of 7 satellites all
  lying at approximately the same distance of $\sim$100 kpc. Both its
  Hess diagram and the CMD clearly show a prominent red giant branch
  as well as the group of the BHB stars. With a luminosity of
  $M_V=-3.5$ and a size of 30\,pc it sits within the uncertain
  territory where extended star cluster and faint dwarf galaxies start
  to overlap. We class it as an "ultra-faint satellite".

\item Pictoris~1 (see Figure~\ref{fig:pictoris1}). Another high
  significance ultra-faint satellite located at $\sim$ 115 kpc and
  projected near the LMC on the sky. The CMD shows a clear RGB and a
  few stars on the HB. Its size is $\sim $ 30\,pc and its luminosity is
  $M_V =-3.2$.

\item Phoenix~2 (see Figure~\ref{fig:phoenix2}). An ultra-faint
  satellite at $\sim$ 80 kpc with the half-light radius of 26\,pc and
  luminosity of $M_V=-2.8$. Apart from a clear MSTO and RGB, one or
  two BHB stars, there is potentially a large number of BS stars
  extending to g$-$r$<$0, r$\sim$22.

\item Indus~1 (see Figure~\ref{fig:indus1}). Similar to the bulk of
  our sample, this ultra-faint satellite is at 100\,kpc with a
  half-light radius of 36\,pc and the luminosity of
  $M_V=-3.5$. Interestingly, Indus~1 is conspicuous in the DES
  false-color image (see Figure~\ref{fig:indus1_eridanus3_color}).

  At the time of the submission of this manuscript, the discovery of
  the same satellite was announced in \citet{kim15} where the object's
  size and luminosity are quoted to be significantly lower than the
  values we obtain. Using imaging data much deeper than available to
  us, \citet{kim15} report the detection of a pronounced mass
  segregation, which (if true) would argue strongly in favour of the
  globular cluster nature of the object. Moreover, the disagreement in
  the inferred total luminosity can be explained by the difference in
  depth between the two datasets in the presence of a depletion at the
  low-mass end of the mass function. With regards to the extent of the
  object, we note that it is possible that our size determination has
  been biased high by blending in the central parts of the
  object. However, the small size reported by \citet{kim15} could, at
  least in part, be the consequence of the choice not to fit the
  stellar density at lager radius, as it is classified as
  ``extra-tidal'' by the authors.

\item Grus~1 (see Figure~\ref{fig:grus1}). The satellite is at
  120\,kpc, has a luminosity of $-3.4$ and a size of 60\,pc. Given its
  luminosity and size, the satellite should perhaps be classified as a
  dwarf galaxy. However, care should be taken when interpreting the
  measured structural parameters, as the object sits right at the CCD
  chip gap.

\item Eridanus~3 (see Figure~\ref{fig:eridanus3}). The satellite is at
  90\,kpc with the smallest half-light radius in the sample, of only
  13\,pc and the luminosity of $M_V=-2.0$. Eridanus~3 is also easily
  noticeable in the DES image (right panel of
  Figure~\ref{fig:indus1_eridanus3_color}).

\item Tucana~2 (see Figure~\ref{fig:tucana2}). This satellite can
  perhaps be classified as an ultra-faint dwarf galaxy with less
  ambiguity than many others in the sample, due to its more
  substantial luminosity of $M_V=-4.4$ and the large size of 160\,pc. The extent of the object is immediately obvious from the map of
  selected stars (second panel in the top row). The CMD demonstrates a
  strong MSTO, a sub-giant branch. { A few blue objects with colors consistent with the 
  BHB stars are probably foreground contaminants as their magnitude contradicts the distance modulus implied
  by the subgiant branch.} The estimated distance is $\sim$ 60\,kpc

\end{itemize}

\subsection{Size-luminosity relation}
\label{sec:mv_rh}

Figure~\ref{fig:rh_mv} shows the locations of the discovered
satellites on the plane of half-light radius ($1.67 a_{maj}
\sqrt{1-e})$ and intrinsic luminosity ($M_V$). Together with the
objects in our sample (red filled circles), several other stellar
systems are displayed. These include globulars clusters (GCs, black
dots), MW dwarf galaxies (black unfilled circles), M31 dwarf galaxies
(black empty triangles), as well as extended objects with half-light
radius less than 100 pc from the catalogue of \citet{brodie11}. Let us
start by pointing out that all groups of objects in this compilation
are affected by selection effects. Most importantly, the
detectablility of all objects is dependent on their surface
brightness. For objects detectable as resolved stellar systems in
surveys similar to the SDSS, the parameter space below the
surface-brightness limit of $\mu \sim 31$ mag deg$^{-2} $
\citep{koposov08} is unexplored (see dashed lines of constant
surface-brightness in the Figure). Additionally, there are constraints
set by blending for star clusters, which deteriorate with increasing
distance.

Notwithstanding various selection effects, star clusters and dwarf
galaxies seem to occupy distinct regions of the size-luminosity
plane. Brighter than $M_V=-6$, star clusters seem to rarely exceed
half-light radius $r_h\sim60$\,pc, while no dwarf galaxy is smaller than
100\,pc. In other words, there seems to be a size gap between star
clusters and dwarf galaxies.  At fainter magnitudes, the picture is a
lot more muddled.  Not only are there objects firmly classified as
dwarf galaxies with size under 100\,pc (for example Pisces II and Canes
Venatici II), but also several ultra-faint satellites, namely Segue 1,
Segue 2 and Bootes 2 have sizes and luminosities comparable to faint
extended globular clusters such as Palomar 4 and Palomar 14. What sets
Segue 1,2 and Bootes 2 apart from similar looking star clusters are
their chemical abundances. All three objects appear to have large
metallicity spreads \citep[e.g.][]{frebel14, kirby13, koch14},
previously unnoticed in star clusters.

\begin{figure*}
\begin{center}
\includegraphics[width=0.95\textwidth]{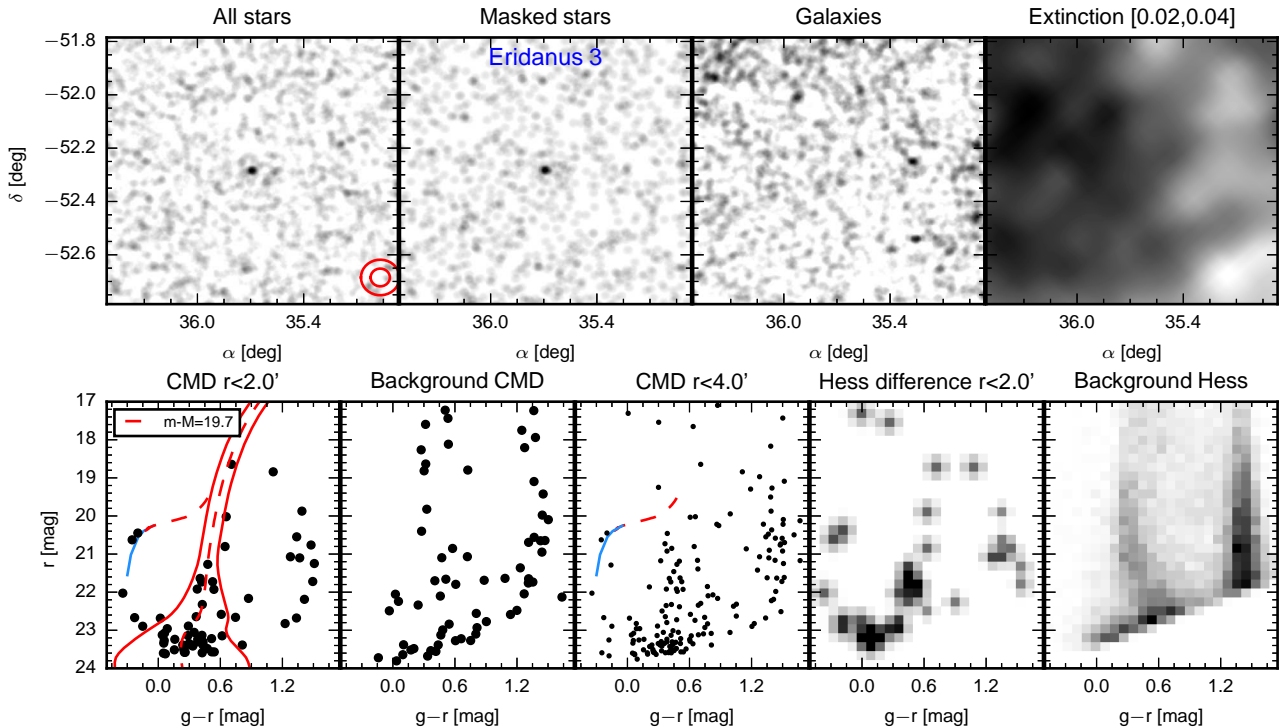}
\end{center}
\caption{Eridanus 3. See Figure~\ref{fig:reticulum2} for detailed
  description of the panels. A very compact satellite that falls in
  the distance range shared by Horologium 1, Pictoris 1, Phoenix 2,
  Indus 1 and Grus 1. The satellite's CMD sequence is rather
  sparse. There are hints of a turn-off, BS stars and an RGB. Despite
  its less convincing CMD, Eridanus 3 is a genuine satellite as can be
  seen from the DES image cut-out shown in
  Figure~\ref{fig:indus1_eridanus3_color}.}
\label{fig:eridanus3}
\end{figure*}
\begin{figure*}[t]
\begin{center}
\includegraphics[width=0.95\textwidth]{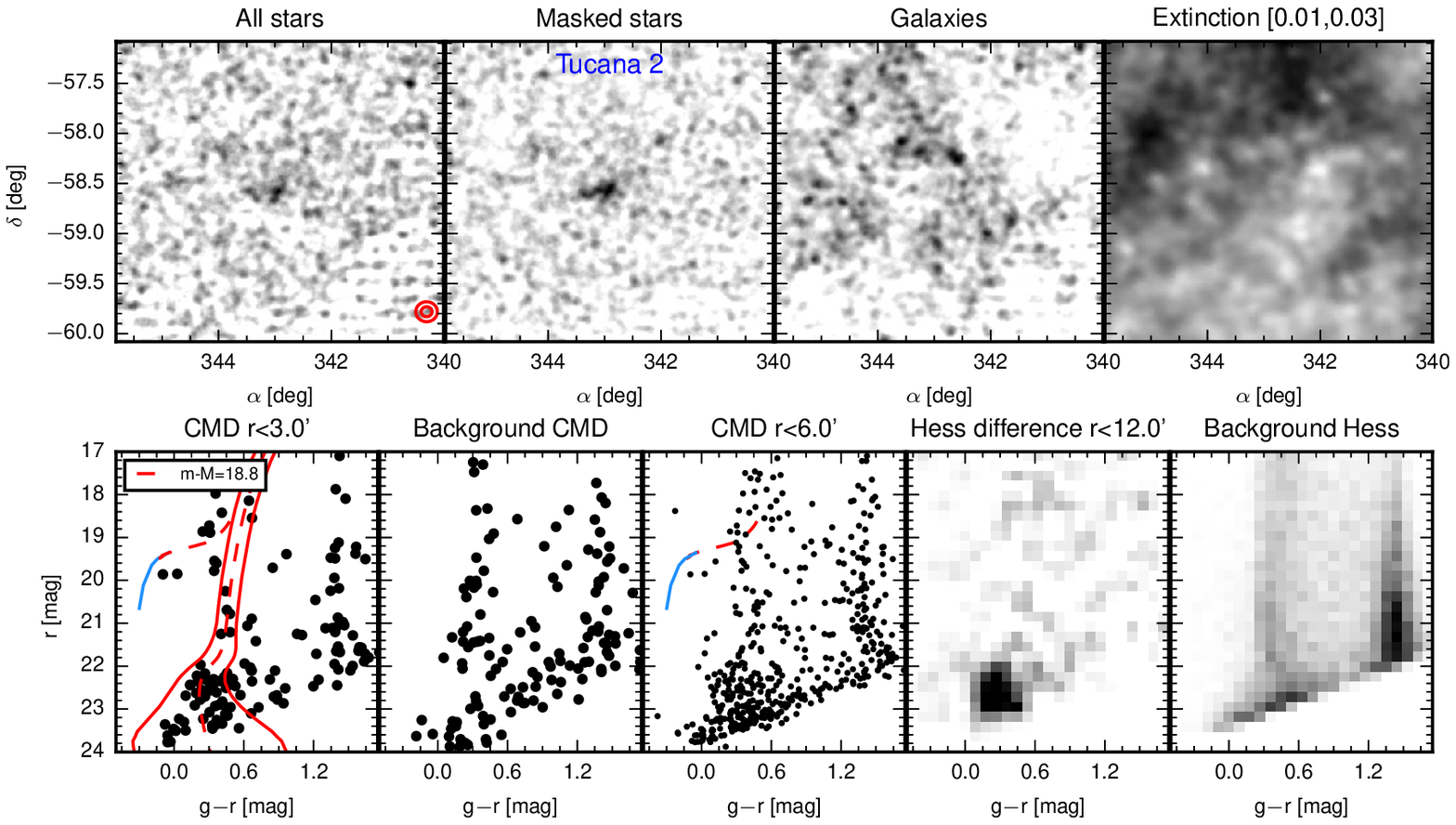}
\end{center}
\caption{Tucana2. See Figure~\ref{fig:reticulum2} for detailed
  description of the panels. Tucana 2 is the closest of the group of 7
  satellites at around 60 kpc. It is considerably fluffier in
  appearance which explains its lower significance. The strongest
  feature in the CMD is the MSTO.}
\label{fig:tucana2}
\end{figure*}

Guided by the approximate shapes of the distribution of star clusters
and dwarf galaxies in size-luminosity plane, we can conclude the
following. Using the measured morphological parameters only, out of
the 9 satellites, only three, namely Eridanus 2, Tucana 2 and Grus 1
can be classified as dwarf galaxies unambiguously, without having to
resort to follow-up spectroscopy. We classify the remaining six,
i.e. Eridanus 3, Phoenix 2, Horologium 1, Pictoris 1, Reticulum 2 and
Indus 1 simply as ultra-faint satellites with the hope that their true
nature is revealed through detailed abundance studies in the near
future.

It is interesting that in the size luminosity plane the ambigiously
classified objects in our sample seem to join the sequence of objects
observed in other galaxies labelled as faint-fuzzies (FF) or Extended
Clusters (EC) \citep[see][]{larsen01,brodie11,forbes13}.  These
objects have sizes of tens of parsecs and luminosities of
$M_V\gtrsim-5$. While our objects are significantly fainter, one
should take into account that most HST surveys of other galaxies are
incomplete for objects with $M_V\lesssim-5$. The possible connection
between FF/EC objects and the ultra-faint satellites is particularly
intriguing given the likely association between some of our objects
and the Magellanic Clouds (see Section~\ref{sec:spatial}).

Strikingly, while most of the new satellites announced here are rather
faint, none of them lies particularly close to (with the exception of
Tucana~2), or indeed below, the nominal surface brightness limit of
$\mu\sim 31$.  This means that these satellites are not very different
from those discovered by Sloan (only somewhat more distant at fixed luminosity), 
and, therefore, significant fraction, but not all of them (as the objects lie very close to the luminosity vs distance limits of 
\citet{koposov08} and \citet{walsh09}) would have very likely been identified 
by the SDSS and the VST ATLAS had they fallen within the footprints of these surveys. 
The relatively high surface
brightness values of the DES satellites must imply that we have not
tapped into a large supply of nearly invisible dwarfs predicted by
some of the semi-analytic models. Instead, the discovery of a large
number of satellites from a small area as compared to the previously
observed sky suggests a peculiar spatial distribution of these
objects. This is the focus of the next Section.

\begin{figure*}
\includegraphics[]{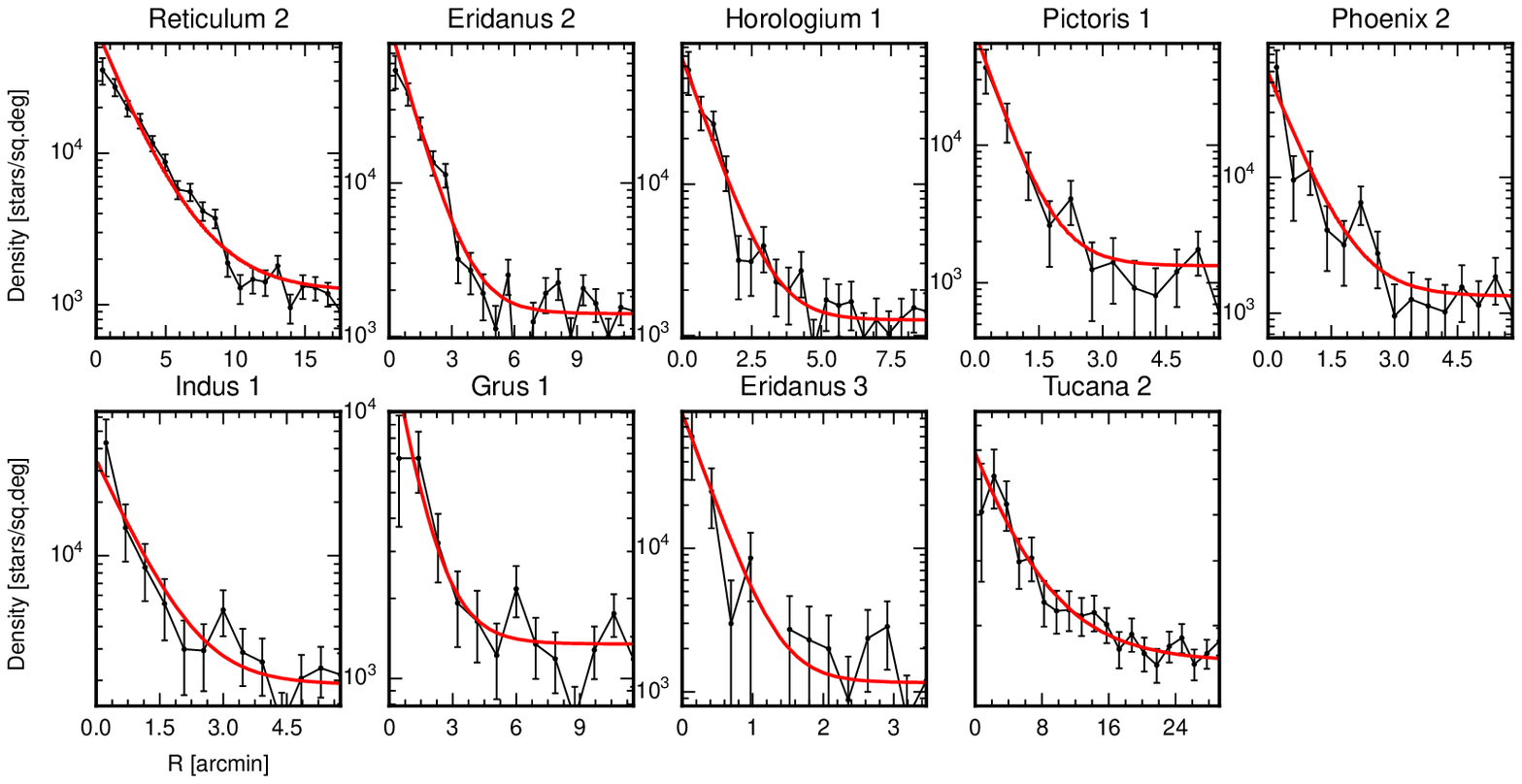}
\caption{Azimuthally averaged 1D density profiles of the detected
  satellites (black curves) with the associated uncertainties. Red
  curves represent the best-fit exponential density profiles. Note that the
actual fits has been done using unbinned 2D data.}
\label{fig:profiles}
\end{figure*}
\begin{figure*}
\begin{center}
\includegraphics[width=0.45\textwidth]{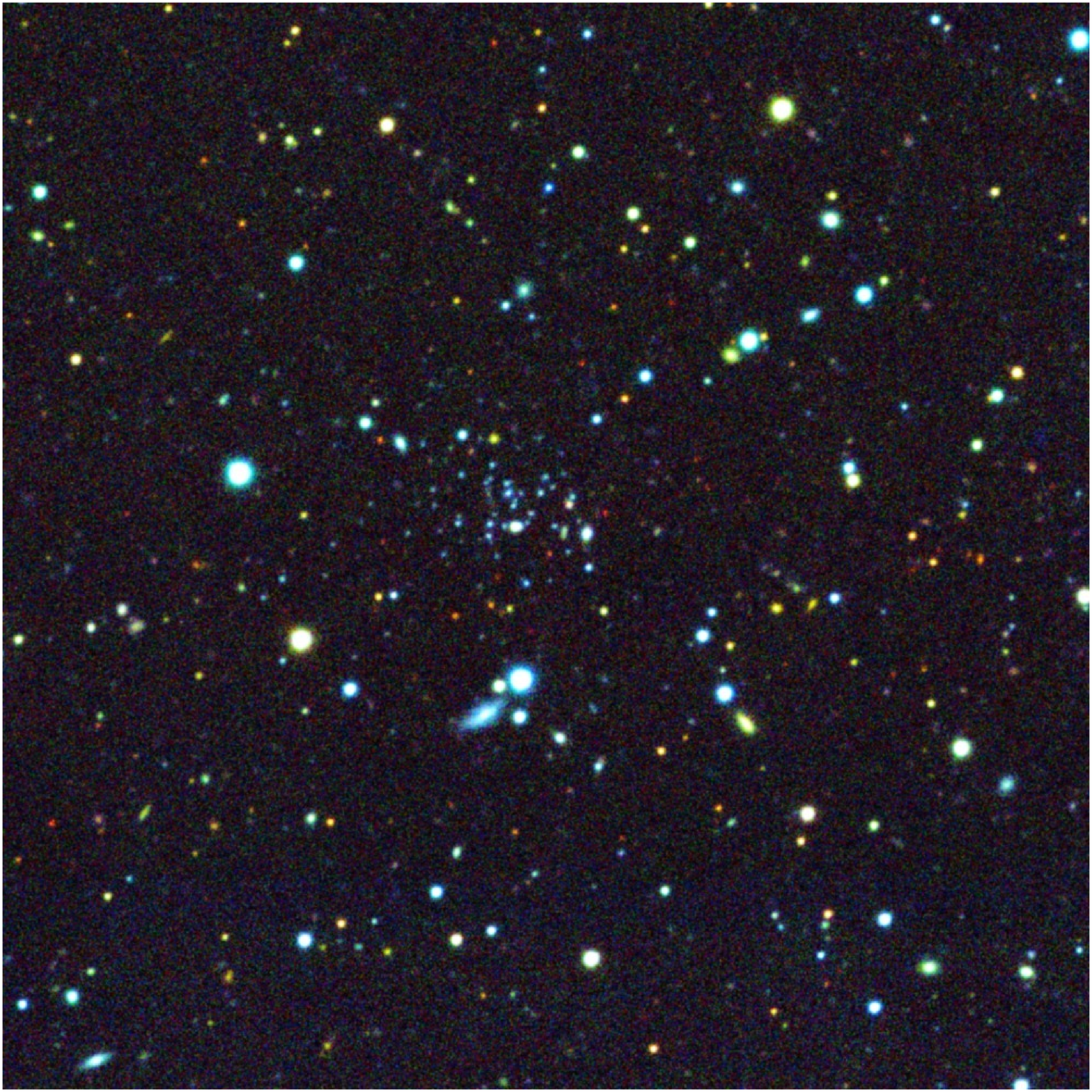}
\includegraphics[width=0.45\textwidth]{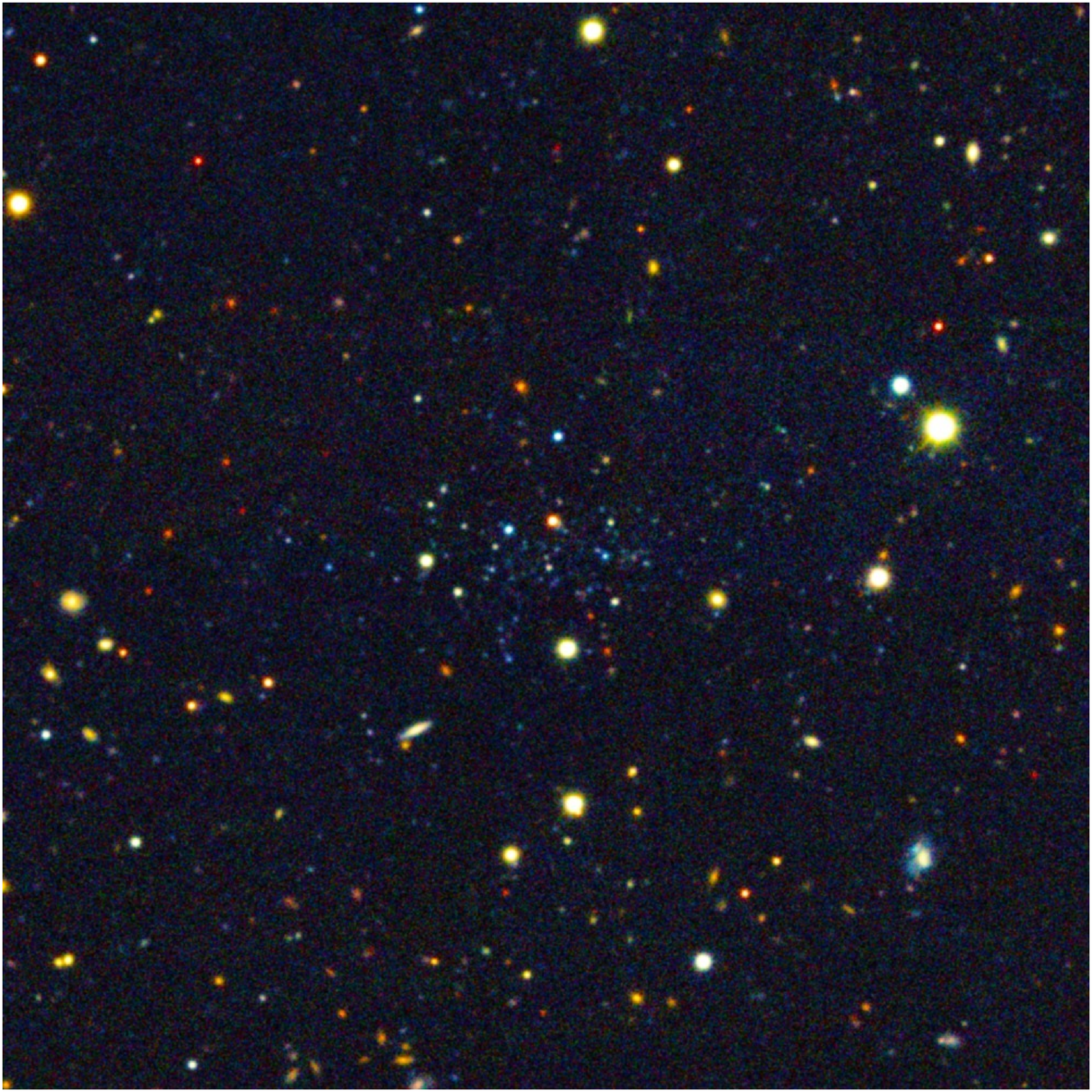}
\end{center}
\caption{DES false-colour composite images of the two compact
  satellites, Indus~1 and Eridanus~3. Each image is 
4.3$\arcmin\times$4.3$\arcmin$. The R,G,B channels correspond to 
the $i$, $r$, $g$ 
bands.}
\label{fig:indus1_eridanus3_color}
\end{figure*}
\begin{figure*}
\begin{center}
\includegraphics[width=0.45\textwidth]{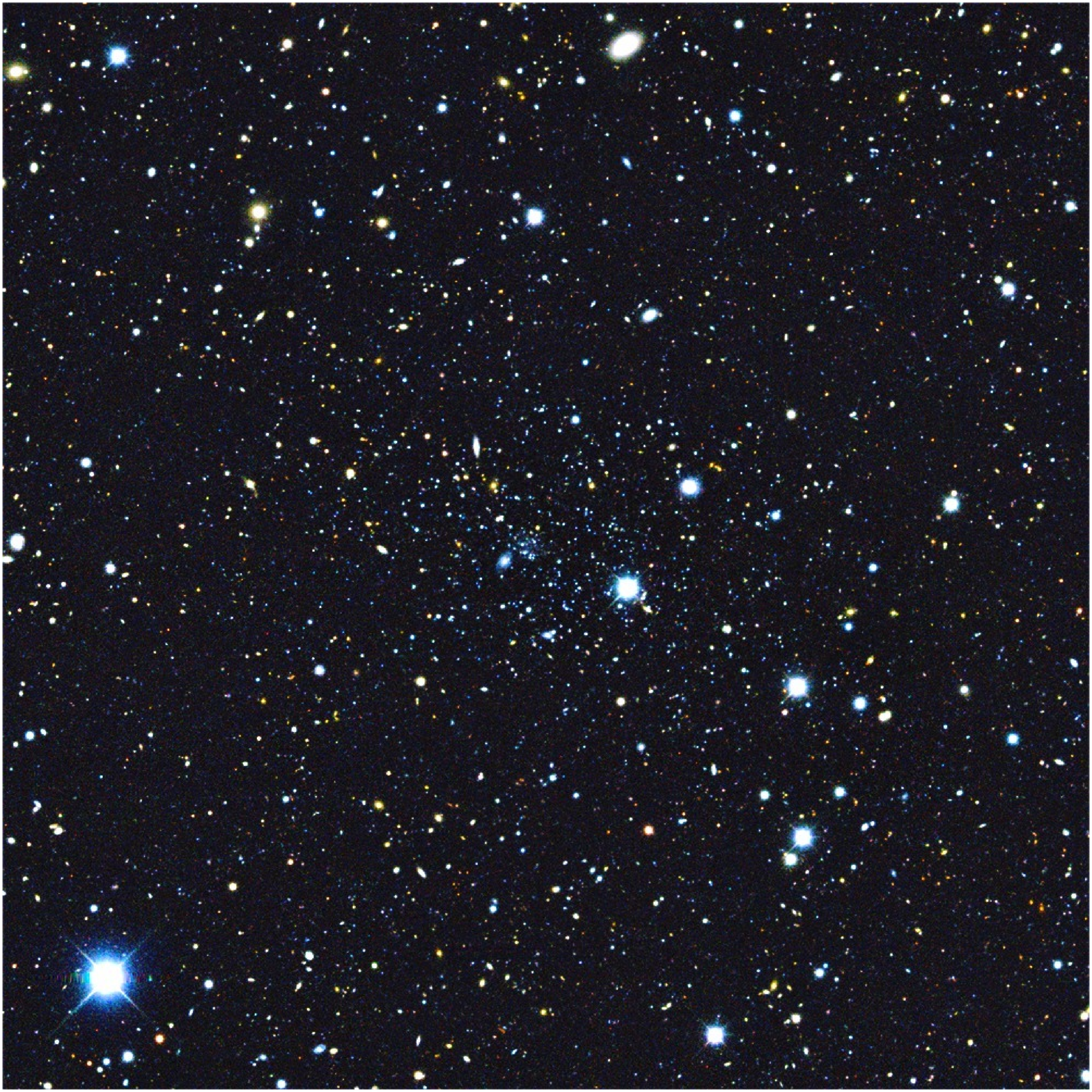}
\includegraphics[width=0.45\textwidth]{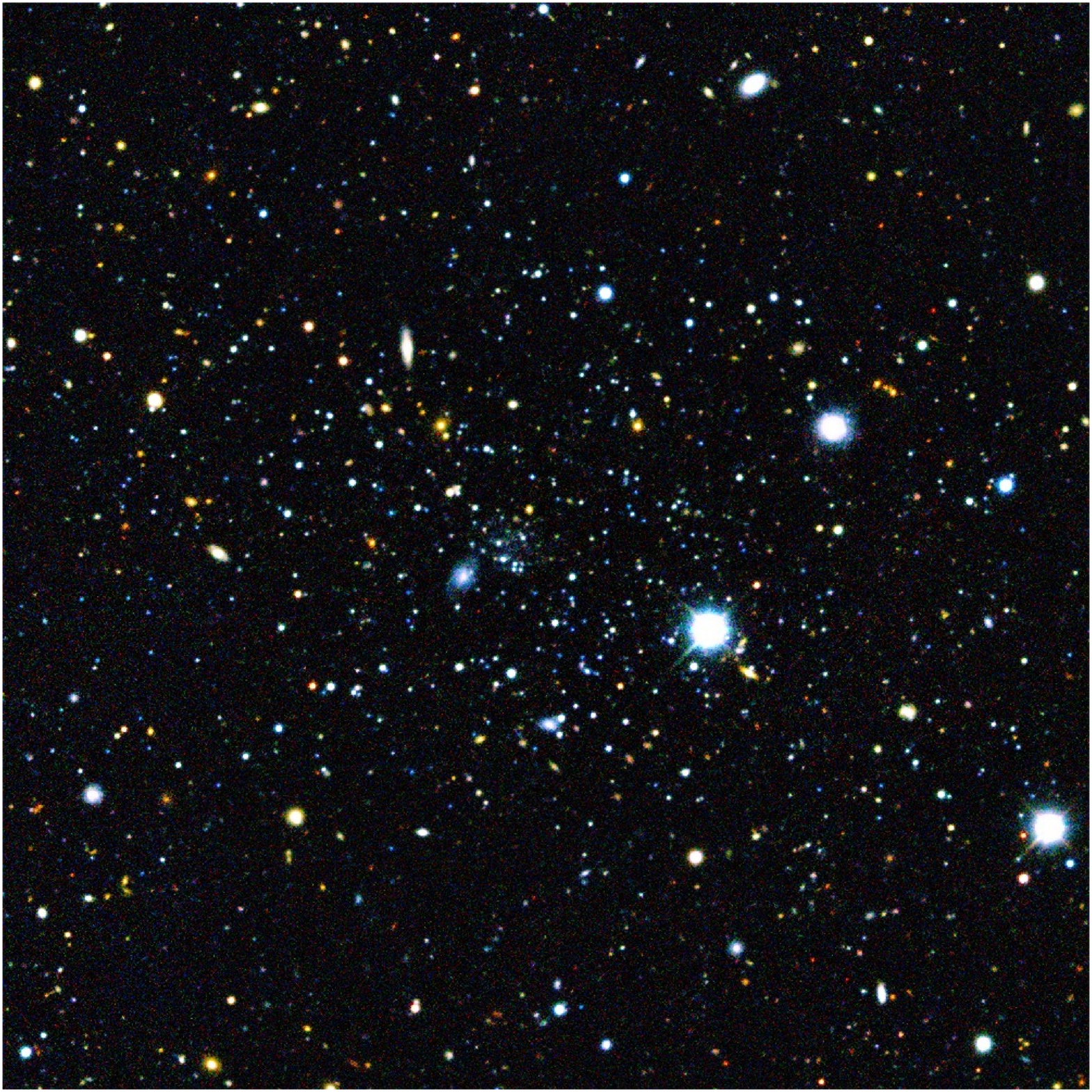}
\end{center}
\caption{DES false-colour composite image of the Eridanus 2 dwarf
  galaxy. {\it Left}: 13$\arcmin\times$13$\arcmin$ field around the galaxy, which
  is visible as a clear concentration of blue stars as well as the low
  surface brightness emission due to unresolved stars. {\it Right}:
  The central regions (6.5$\arcmin\times$6.5$\arcmin$) of the Eridanus~2
  dwarf galaxy. The very low surface brightness round spot is clearly
  visible near the center and is possibly a globular cluster
  associated with the galaxy. 
  A brighter elongated blue object below is more likely to be a background
source.
}
\label{fig:eridanus2_color}
\end{figure*}

\begin{figure}
\includegraphics{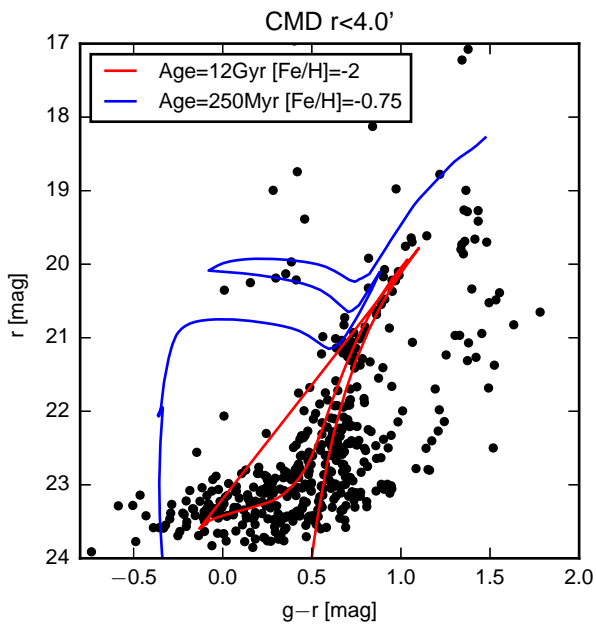}
\caption{Evidence of young stellar populations in the color-magnitude
  diagram of the Eridanus~2 dwarf galaxy. As well as the old
  metal-poor stars (12 Gyr, ${\rm [Fe/H]}=-2$ isochrone, red curve), the dwarf
  galaxy seems to show a significant Red Clump and a handful of Blue
  Loop stars indicative of recent star-formation as revealed by the
  metal-rich 250 Myr isochrone (blue curve).}
\label{fig:iso_eridanus2}
\end{figure}
\begin{figure*}
\includegraphics[width=0.99\textwidth]{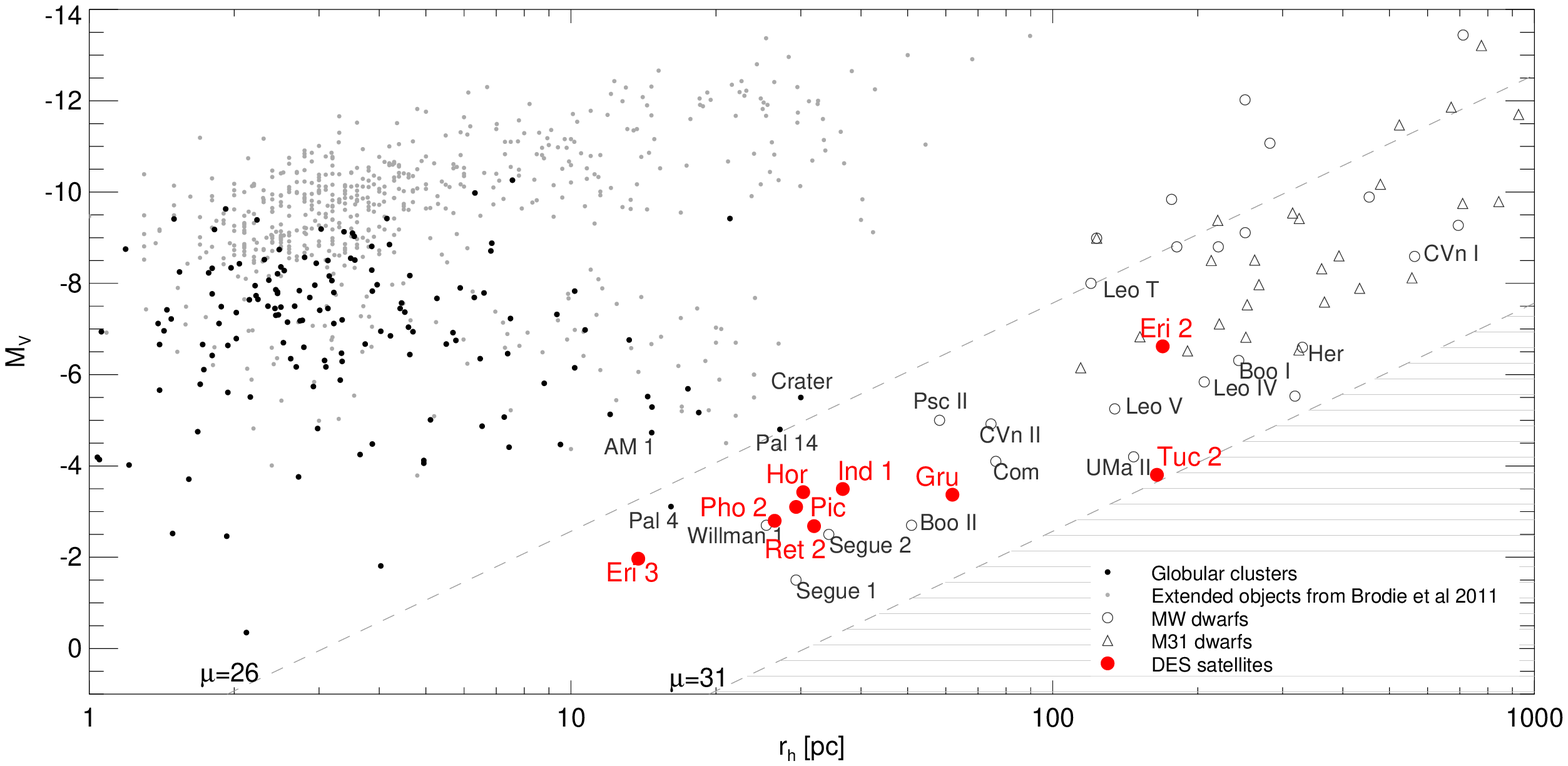}
\caption{Luminosity as a function of half-light radius for star
  clusters and galaxies. MW globular clusters (black dots,
  \citet{francis14}), extended objects with half-light radius less
  than 100 pc (grey dots, from \citet{brodie11}), MW dwarf satellites
  (black circles, \citet{mcconnachie12}), M31 dwarf satellites (black
  triangles, \citet{mcconnachie12}) and the DES satellites (red filled
  circles) are shown. Dashed grey boundaries are curves of constant
  surface brightness $\mu=26$ mag deg$^{-2}$ and $\mu=31$\,mag
  deg$^{-2}$. Grey hatched region is the portion of the parameter
  space currently unaccessible to surveys like SDSS and DES. Note that
  the DES satellites do not lie too close to the low
  surface-brightness limit and, therefore, could have been detected by
  SDSS. Most of the DES objects (apart from Eri 2, Tuc 2 and Gru 1)
  lie in the region right between the extended globular clusters and
  faint dwarf galaxies. Thus, their classification remains ambiguous
  with the data in hand.}
\label{fig:rh_mv}
\end{figure*}

\section{Spatial distribution}
\label{sec:spatial}

\begin{figure}
\includegraphics[width=0.48\textwidth]{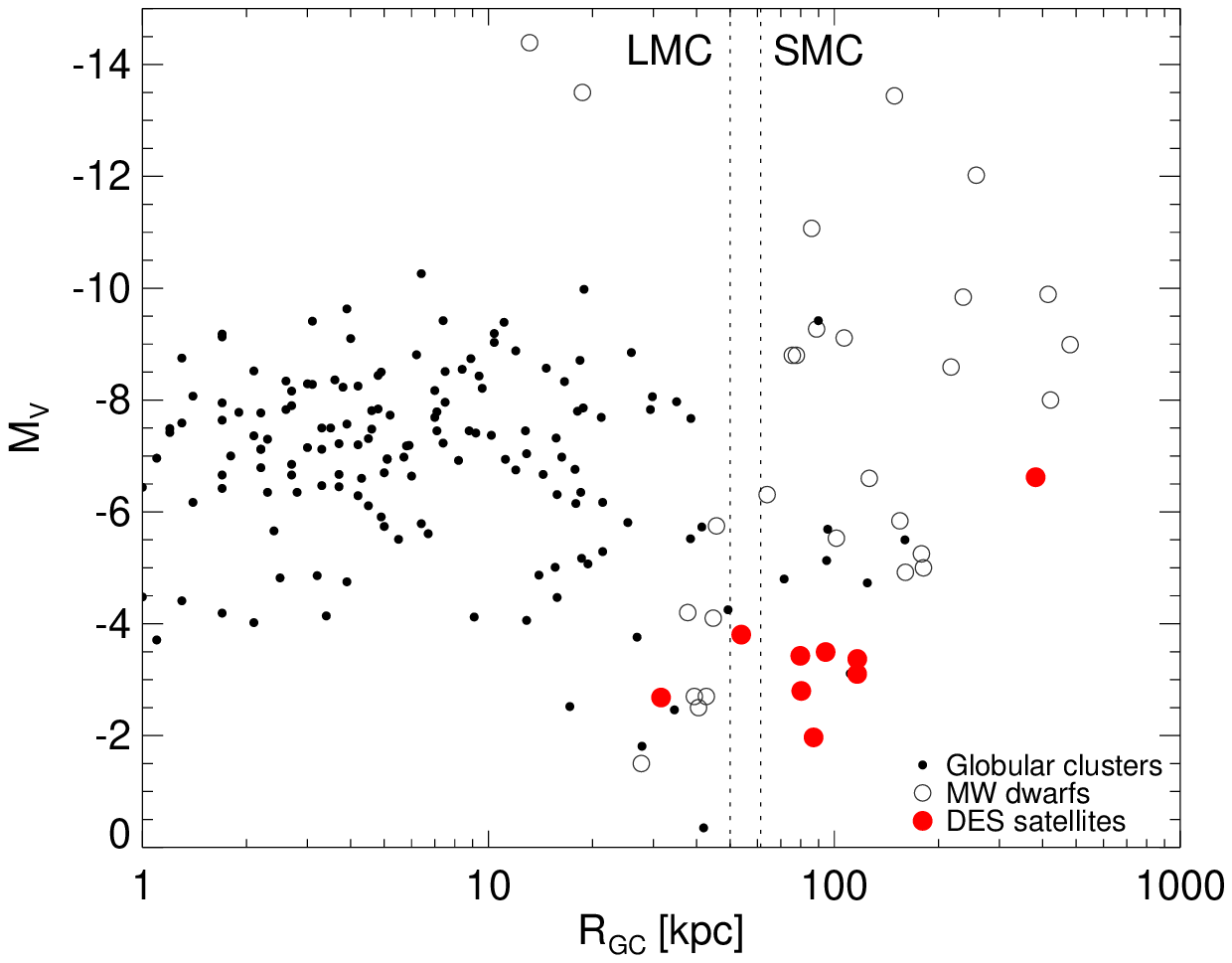}
\caption{Luminosity as a function of Galactocentric distance for
  globular clusters (black dots), MW dwarf galaxies (black unfilled
  circles) and the DES satellites (red filled circled). Note a large
  group of the satellites congregating just outside the SMC orbit
  (vertical dashed line).}
\label{fig:mv_dist}
\end{figure}

Let us start by inspecting the overall distance distribution of the
DES satellites. Figure~\ref{fig:mv_dist} gives the object's luminosity
as a function of Galacto-centric radius $R_{GC}$. It is immediately
obvious that i) our satellites are some of the faintest of the distant
MW satellites known, and ii) a large faction of them cluster around
the distance $R_{GC}\sim100$\,kpc. In other words, the satellites as a
group lie not far from the LMC ($R_{GC}\sim50$\,kpc) and just behind the
orbit of the Small Magellanic Cloud (SMC, $R_{GC}\sim70$\,kpc).

\begin{figure}
\includegraphics[width=0.49\textwidth]{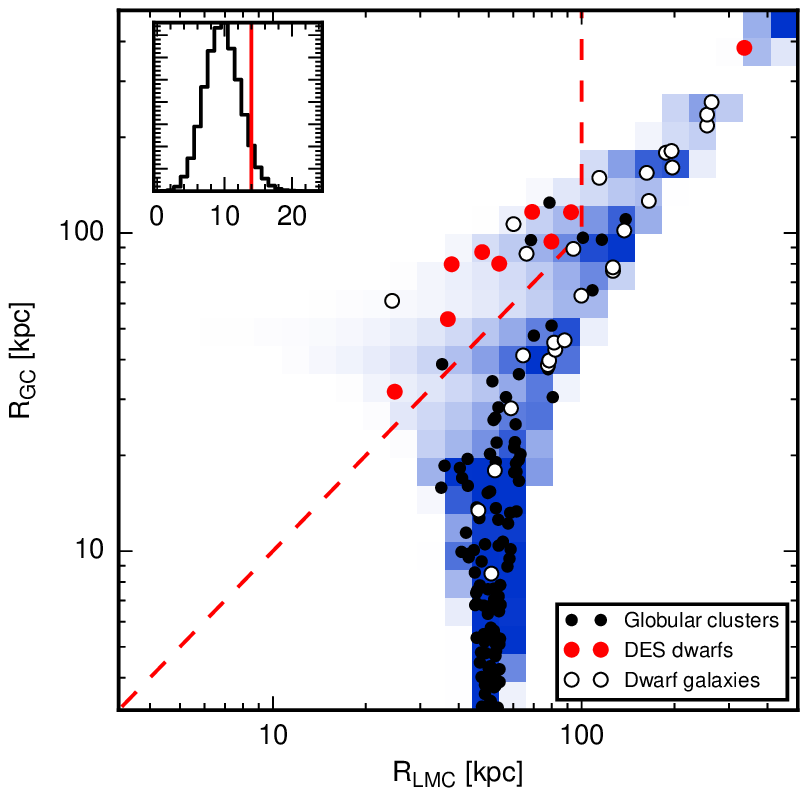}
\caption{Distance to the Galactic center as a function of the distance
  to the LMC. The symbols follow the same scheme as in
  Figure~\ref{fig:mv_dist}. The blue-white 2D histogram in the
  background gives the expected density of objects assuming isothropic
  distribution on the sky (see text for more details). The red dashed
  line defines the "zone of influence" of LMC. The inset shows the
  distribution of the number of objects inside the zone of influence
  as predicted by reshuffling the position vectors of the known
  satellites. The red solid line in the inset shows the actual number of
  satellites in the "zone of influence" of LMC. We conclude that the
  detected objects constitute an overdensity around the LMC with a
  significance of 94\%.}
\label{fig:lmc_rgc}
\end{figure}

A more direct test of a possible association with the Magellanic
Clouds is shown in Figure~\ref{fig:lmc_rgc}. Here we plot the distance
to the LMC against the distance to the Galactic centre for the DES
satellites as well the known MW dwarf galaxies \citep{mcconnachie12}
and globular clusters \citep{francis14}. In this space, the objects
that lie above the one-to-one diagonal line of the plot are closer to
the LMC than to the Galactic centre. Note that all our objects lie
above this demarcation line. While surprising at first, this is in
fact expected due to the particular sky coverage of the DES survey
(see Figure~\ref{fig:colormap}). However, even if the bunching around
the Magellanic Clouds is set by the footprint, the total number of the
satellites uncovered in such small area is alarming, especially when
compared to the Southern portions of the SDSS, or VST ATLAS.

To test whether there is a statistically significant overdensity of
satellites in the vicinity of the LMC we carry out the following
calculation. The 3D position vector of each MW satellite, including
the newly discovered ones, but excluding the LMC itself, is rotated
randomly 50000 times with respect to the Galactic center (GC) while keeping
fixed its Galacto-centric radius. After each rotation, the distance to
the LMC $\rm \widetilde{R}_{LMC}$ is computed and stored. The
blue/white background density in Figure~\ref{fig:lmc_rgc} shows the
average expected density of objects in the ($\rm \widetilde{R}_{LMC}$,
${\rm R_{GC}}$) space from the random 3D position reshuffles described
above. Under the assumption of an isotropic distribution of
satellites in the Galaxy, our procedure would yield a reasonable
estimate of the expected number of satellites in each portion of the
($\rm {R}_{LMC}$, $\rm R_{GC}$) plane. Darker shades of blue show
areas where more objects are expected, while pale blue and white
pixels correspond to less populated corners of the parameter space.

According to the simulation described above, the entire group of the
satellites announced in this paper (excluding the distant Eridanus~2)
is located in the under-populated area of the ($R_{LMC}$, $R_{GC}$)
space, as indicated by the pale blue background density. This can be
interpreted as a detection of an over-density of satellites around the
LMC. To quantify the significance of this detection we perform the
following frequentist hypothesis test.
We carve the LMC-dominated area in the ($\rm R_{LMC}, R_{GC}$) space:
$\rm R_{LMC}<R_{GC}$ and $R_{LMC}<100$\,kpc.
Then the total number of satellites in the LMC-dominated area
(excluding LMC itself) is computed.  The actual observed number
($N=14$) is compared to the number of objects in the marked region as
given by the random reshuffles of the 3D position vectors.  The
distribution of the possible number of satellites in the marked area
is shown in the inset of the Figure, and the observed number is given
by the red vertical line. The distribution obtained assuming isotropy
is narrow and peaks around $N=10$, indicating the presence of a
satellite overdensity in the DES data. However, the formal
significance level of this result is not very high at $\sim$
94\%\footnote{The boundary in the $\rm R_{LMC}$ vs $\rm R_{GC}$ space
  used for the test is chosen to be as simple as possible to avoid
  artificially inflating the p-value.} while the estimated excess is
also quite modest, i.e. of order of 4 objects.

\begin{figure}
\includegraphics[width=0.49\textwidth]{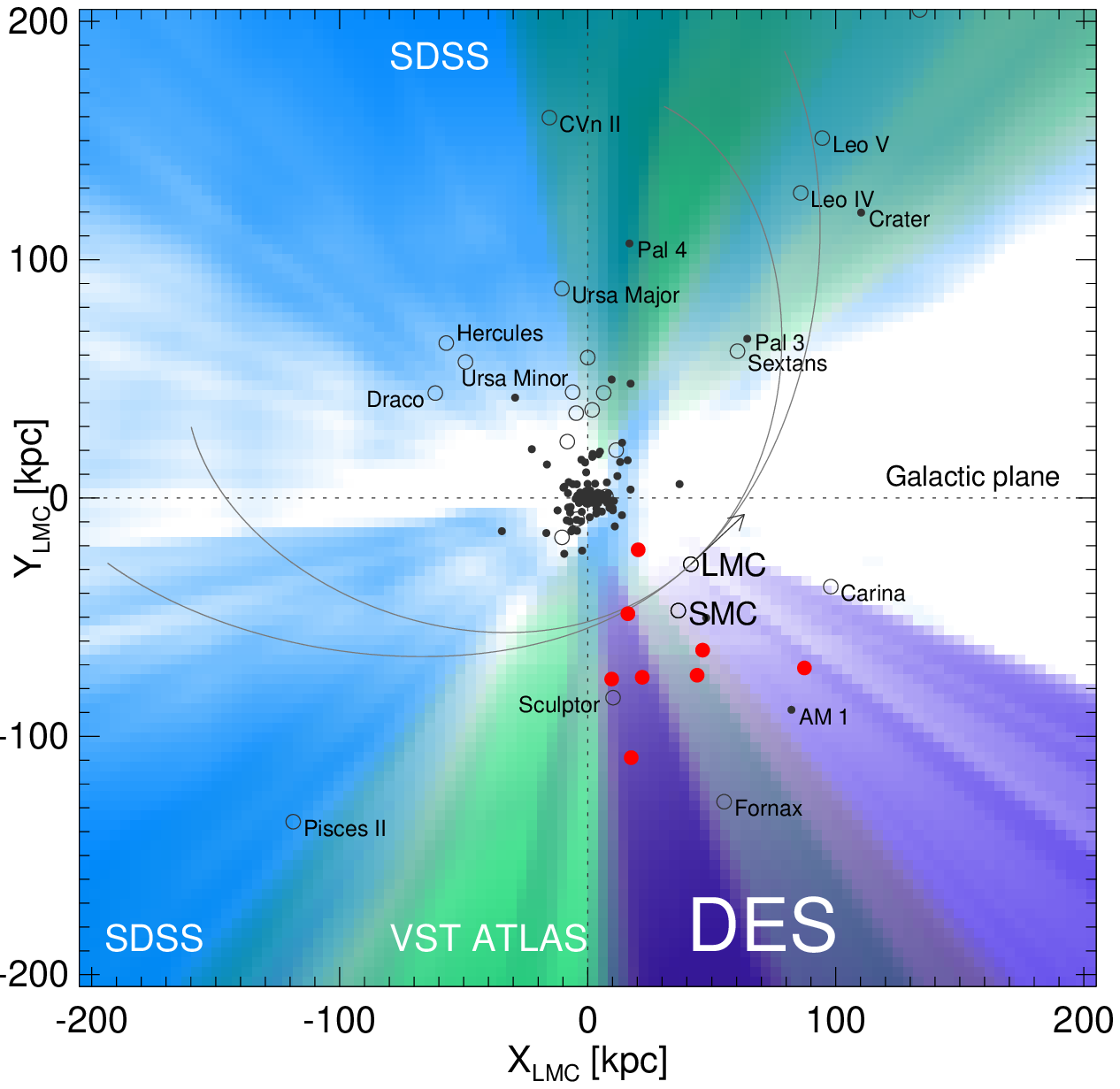}
\caption{Distribution of the MW satellites in the orbital plane of the
  LMC. The symbols are assigned following the convention described in
  Figure~\ref{fig:mv_dist}. Coloured 2d histograms reflect the
  projection of the surveyed volume of SDSS (blue), VST ATLAS (green)
  and DES (lilac), with darker shades indicating nearly complete
  coverage of the volume out to 250 kpc. Arrow shows the direction of
  the LMC's motion as measured by \citet{kallivayalil13}. Four LMC
  orbits are shown: two integrated forward in time and two backward in
  time for two different MW potential, one with a DM halo with
  concentration $c=10$ and one with $c=24$.}
\label{fig:lmc_plane}
\end{figure}

\begin{figure}
\includegraphics[width=0.49\textwidth]{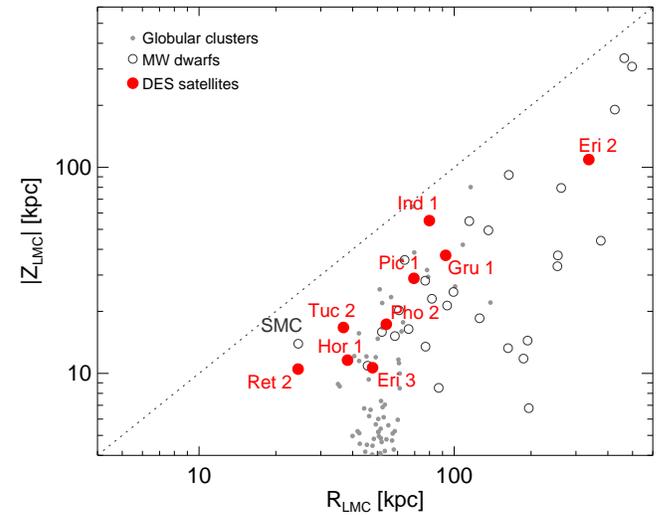}
\caption{Distance from the LMC orbital plane $|Z_{LMC}|$ as a function
  of the distance from the center of the LMC. Symbols are assigned as
  before, albeit the MW globular clusters are grey to aid
  clarity. Only three 3 of 9 new satellites show alignment with the
  LMC's orbital plane. These are Reticulum 2, Horologium 1 and
  Eridanus 3.}
\label{fig:lmc_zlmc}
\end{figure}

Having established a possible connection between the Magellanic Clouds
and the DES satellites, we study the distribution of the new objects
in the plane of the LMC's orbit. Figure~\ref{fig:lmc_plane} shows the
MW satellites (GCs, dwarfs and DES satellites) projected onto the
plane defined by the vector of the LMC's velocity as given in
\citet{kallivayalil13}\footnote{ The transformation from the
Galactocentric coordinates (X,Y,Z) (X is positive towards the Galactic anticenter) 
to $X_{LMC},Y_{LMC},Z_{LMC}$ shown in Figures~\ref{fig:lmc_plane} and
\ref{fig:lmc_zlmc} is defined
by
$X_{LMC}=     0.10490391 \, X -0.99448236\,Y +     0\,Z$;   
$Y_{LMC}=     0.14676983 \, X +   0.015482154\,Y +      0.98904950\,Z$;
$Z_{LMC}=  -0.98359229 \,X    -0.10375516 \,Y +     0.14758415\,Z$.
}. We also show the forward- and
backward-integrated orbits of the LMC in two MW potentials with
different DM halo concentrations ($c=10$ and $c=25$). Coloured regions
correspond to the fraction of the Galactic volume covered by the SDSS,
VST ATLAS and DES surveys as projected onto the LMC's orbital
plane. The three slightly over-lapping surveys have together covered a
large portion of the sky. However, some lacunae still remain, most
notably directly in front of the LMC-SMC pair as judged by the vector
of the LMC motion (black arrow). As expected, the DES satellites bunch
tightly outside the SMC's orbit. However, while the objects have
similar in-plane coordinates, not all of our satellites actually lie
close to the orbital plane of the LMC. The distance from the LMC's
orbital plane is shown in Figure~\ref{fig:lmc_zlmc}. Only, Reticulum
2, Horologium 1 and Eridanus 3 have small heights above the plane
$|z_{LMC}| < 10$ kpc. The distribution of the other 6 objects do not
show any strong alignment with the LMC's orbit.

\begin{figure*}
\begin{center}
\includegraphics[width=0.95\textwidth]{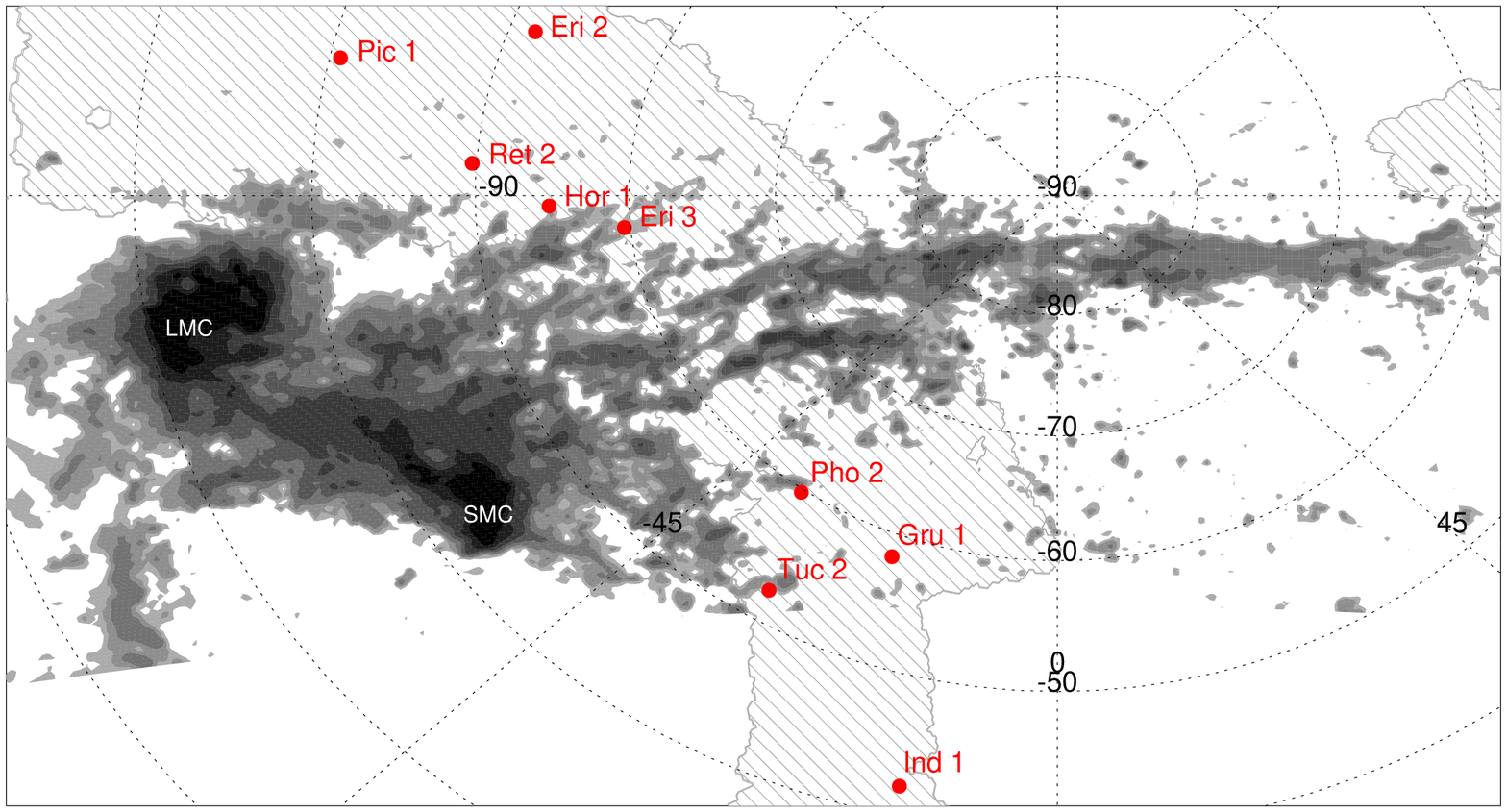}
\end{center}
\caption{Positions of the DES satellites with respect to the
  Magellanic gaseous Stream near the Southern Galactic Pole. The
  Stream's HI column density from \citet{putman03} is shown as filled
  contours, with darker shades corresponding to higher
  densities. Hatched area contour shows the current DES footprint. The
  DES satellites appear to be avoiding regions with high HI column
  density.}
\label{fig:lmc_gas}
\end{figure*}

The accretion of the Magellanic system has left behind a trail of
neutral hydrogen. Figure~\ref{fig:lmc_gas} compares the positions of
the DES satellites to the distribution of HI in the Magellanic Stream
as traced by \citet{putman03}. Curiously, the DES satellites seem to
avoid the high column density portions of the stream.  { The spatially
distinct behavior of the gas and the stars is similar to that seen
around M31~\citep{Le13}.} There is however possible overlap between
Horologium 1, Eridanus 3, Phoenix 2 and Tucana 2 with the high
velocity clouds associated with the Stream. Additionally, the
particular projection used in the Figure emphasizes the fact that
there appears to be two groups: one containing Pictoris 1, Eridanus 2,
Reticulum 2, Horologium 1 and Eridanus 3, all lying on the LMC side of
the Stream. Meanwhile, Phoenix 2, Tucana 2, Grus 1 and Indus 1 can be
found on the SMC side. Interestingly, the three objects that have the
highest probability of being aligned with the orbital plane of the LMC
(see Figure~\ref{fig:lmc_zlmc}), i.e. Reticulum 2, Horologium 1 and
Eridanus 3 are the satellites closest to the Magellanic Stream.


\section{Discussion and Conclusions}

In this paper, we have presented the discovery of 9 new ultra-faint
Milky Way satellites using the data from the Dark Energy Survey. Based
on the morphological properties, 3 of the satellites are definite
dwarf galaxies, while the exact type of the other 6 is more ambiguous.
The nearest object in the sample shows a significant elongation
indicative of the influence of the Galactic tides. The most distant
object discovered appears to be a dwarf galaxy located at the very
edge of the Milky Way halo, around $\sim$ 380\,kpc. This remote dwarf
shows signs of recent star formation and possibly even hosts a faint
globular cluster! The discovery of a large number of MW satellites in
a small area around the LMC and the SMC is suggesting that at least a
fraction of the new objects might have once been a part of the
Magellanic group.

The idea that some of the satellite galaxies of the Milky Way may be
associated with the Magellanic Clouds is quite old. Let us re-visit
the earlier work in the light of the discoveries in this paper.

\citet{lyndenbell76} first suggested the idea of a Greater Magellanic
Galaxy. His paper predates modern ideas on galaxy formation, and so
his picture is rather different from today's concensus. He envisaged
dwarf galaxies condensing from gas torn out of the Greater Magellanic
Galaxy via a tidal encounter with the Milky Way Galaxy. He noted that
Draco and Ursa Minor lie in the Magellanic Plane, almost opposite the
Magellanic Clouds in the Galactocentric sky, and were candidates for
debris from the break-up of the Greater Magellanic Galaxy. This
speculation received some confirmation when \citet{Ly82} showed
that the elongation of Draco and Ursa Minor is along the Magellanic
Stream.  \citet{Ly95} later suggested that
Sculptor and Carina can also be added to the association of galaxies
in the Magellanic Stream, though their membership is more tentative as
their orientations are not along the Stream. Nonetheless, the evidence
is at least suggestive for a linkage between Draco, Ursa Minor and the
Magellanic Clouds.

\citet{odonghia08} argued that the Magellanic Clouds were the
largest members of a group of dwarf galaxies that fell into the Milky
Way halo at a comparatively late time. This was inspired by \citet{Tu06}
observational discovery of a number of nearby
associations of dwarf galaxies just outside the Local Group at
distances of $\sim$ 5 Mpc. These groups have masses $\sim 1-10 \times
10^{11} M_\odot$ and typically contain $\sim 5$ dwarfs comparable to
in mass to the larger dwarf spheroidals, as well as an unknown number
of fainter satellites. The sizes of the dwarf associations are a few
hundred kiloparsecs, and the velocity dispersions a few tens of
kms$^{-1}$. They are bound, but not in dynamical equilibrium as the
crossing time is a sizeable fraction ($\sim 80 \%$) of the Hubble
time. \citet{odonghia08} speculated that the LMC may have been
the largest member of such a dwarf association. When this Magellanic
Group fell into the Local Group, it provided seven of the bright
satellites of the Milky Way (LMC, SMC, Draco, Sex, Sgr, UMi, Leo
II). They argued that subhalos accreted singly may remain dark, whilst
subhalos accreted in groups may light up. This is because the low
virial temperature in the parent halo allows gas to cool and be
re-accreted onto subhalos, though this argument was not corroborated
by hydrodynamical simulations.  Although it is unclear whether the
seven dwarfs proposed by \citet{odonghia08} really all were part
of a Magellanic Group, it is certainly true that group infall occurs
frequently in numerical simulations \citep[e.g.][]{Li08,wetzel15}
 and so the underlying motivation is sound.

{ \citet{sales11} studied the orbits of the analogues of the
  Magellanic Clouds in the Aquarius simulations
  \citep{springel08}. Although their analysis ruled out Draco and UMi
  as possible members of the Magellanic group, they suggested that
  there could possibly still remain a large number of faint satellites
  near the Clouds, in particular if the LMC and the SMC are on their
  first approach of the MW.}

{ There is also evidence of satellite associations around other nearby
galaxies. \citet{Ib13} argued that half of the satellites of M31 may
lie in a thin, extended and rotating plane. Measurements of velocities
of pairs of diametrically opposed satellite galaxies also hint that
planes of satellites may be common in the low redshift
Universe~\citep{Ib14}.  Very recently, \citet{Tu15} has argued that
all but 2 of the 29 galaxies in the Centaurus Group lie in one of two
planes. Satellite galaxies are clearly not isotropically distributed
and correlations are widespread.}
 
This suggests a picture in which some or all of the objects described
in this paper -- together with possibly some of the classical
satellites like Draco and Ursa Minor -- really did belong to a loose
association with the LMC and the SMC as its largest members. The
association would then have a mass of 5 to 10 per cent of the mass of
the Milky Way. Given the present-day distances, which range from 30
kpc (Reticulum 2) to 380 kpc (Eridanus 2) together with the bunching
of seven satellites at 100 kpc, then the overall radius of the
association must have been $\sim 150$ kpc.  These are very comparable
to the masses and radii of the dwarf associations discovered by \citet{Tu06}.

What might the orbit of such a Magellanic Group be?  \citet{kallivayalil13} measured the proper motion of the LMC using a 7 year
baseline. Orbit calculations depend on the uncertain masses of the
Milky Way and the LMC, but the LMC is typically on first infall. Only
if the mass of the Milky Way exceeds $2 \times 10^{12} M_\odot$ and
the LMC is comparatively light ($3 \times 10^{10} M_\odot$) can more
than one pericentric passage occur. First infall solutions are also
attractive as they also solve the problem of the anomalous blueness of
the LMC. \citet{To11} examined a spectroscopic sample of
isolated galaxies in SDSS, and found that bright satellite galaxies
around Milky Way-type hosts are redder than field galaxies with the
same luminsity.  However, the LMC is unusually blue, which is
plausibly a consequence of triggered star formation upon first
infall. If the LMC is on first infall and if the mass of the Milky Way
is $1 \times 10^{12} M_\odot$ \citep[e.g.][]{Gi14}, then the
orbital period is 9-12 Gyr. Increasing the mass of the Milky Way to
$1.5 \times 10^{12} M_\odot$ gives orbital periods of 7-9 Gyr. This
would imply that the Magellanic Group fell into the Milky Way halo at
about a redshift of $z \approx 1$.  

Motivated by the data of \citet{Tu06}, \cite{Bo14}
looked at the survivability of loose associations in the outer parts
of galaxies like the Milky Way. They found that the timescale required
for an association to phase mix away is 10-15 times longer than the
orbital period (see their Figure 2). So, the structure of such a
Magellanic Group can therefore persist for enormous times, given the
already long orbital periods.

One prediction of the idea of a Magellanic Group is that the
satellites might be expected to share the same sense of ciculation as
the LMC and SMC about the Galaxy. However, on reflection, it is
unclear whether this really need be the case.  Modern simulations
indicate that the LMC and SMC have recently come close enough to do
spectacular damage to themselves \citep[e.g.,][]{Be10,Diaz11}
The large-scale gaseous structures of the Magellanic
stream, the off-centered bar of the LMC and the irregular geometry of
the SMC are all now believed to be the results of recent encounters
between the LMC and the SMC, rather than any interaction with the
Milky Way.  As such a violent interaction proceeds, any satellites or
globular clusters of the LMC or SMC could be flung backwards against the
prevailing circulation.

A firmer -- though rather less specific -- prediction is that there
must be more satellites in the vicinity of the Magellanic Clouds. We
have shown that Reticulum 2, Horologium 1, and Eridanus 3 are aligned
with the LMC's orbital plane and form part of the entourage of the
LMC. Similarly, Tucana 2, Phoenix 2 and Grus 1 most likely comprise
part of the entourage of the SMC. However, all the objects in this
paper trail the LMC and SMC. There must also be a counter-population
of satellites that lead the LMC and SMC. Hunting is always easier
once we know where the big game is plentiful. And the happy hunting
grounds for the \lq beasts of the southern wild' are the Magellanic
Clouds, especially in front of the LMC/SMC pair as reckoned by the
LMC motion.

\section{Acknowledgments}

We wish to thank Mike Irwin, Prashin Jethwa, Mary Putman, Josh Peek,
Filippo Fraternali and Gerry Gilmore for their insightful comments
that helped to improve this paper. Also we would like to thank Richard
McMahon for his help.

On 21 February 2015, after the submission to ApJ this manuscript was 
shared with the DES collaboration. 

This research was made possible through the use of the AAVSO
Photometric All-Sky Survey (APASS), funded by the Robert Martin Ayers
Sciences Fund. This work was performed using the Darwin Supercomputer
of the University of Cambridge High Performance Computing Service
(http://www.hpc.cam.ac.uk/), provided by Dell Inc. using Strategic
Research Infrastructure Funding from the Higher Education Funding
Council for England and funding from the Science and Technology
Facilities Council.  This research has made use of the SIMBAD
database, operated at CDS, Strasbourg, France. Based on observations
obtained with MegaPrime/MegaCam, a joint project of CFHT and CEA/IRFU,
at the Canada-France-Hawaii Telescope (CFHT) which is operated by the
National Research Council (NRC) of Canada, the Institut National des
Science de l'Univers of the Centre National de la Recherche
Scientifique (CNRS) of France, and the University of Hawaii. This work
is based in part on data products produced at Terapix available at the
Canadian Astronomy Data Centre as part of the Canada-France-Hawaii
Telescope Legacy Survey, a collaborative project of NRC and CNRS.

Funding for the DES Projects has been provided by the U.S. Department
of Energy, the U.S. National Science Foundation, the Ministry of
Science and Education of Spain, the Science and Technology Facilities
Council of the United Kingdom, the Higher Education Funding Council
for England, the National Center for Supercomputing Applications at
the University of Illinois at Urbana-Champaign, the Kavli Institute of
Cosmological Physics at the University of Chicago, Financiadora de
Estudos e Projetos, Fundação Carlos Chagas Filho de Amparo à
Pesquisa do Estado do Rio de Janeiro, Conselho Nacional de
Desenvolvimento Científico e Tecnológico and the Ministério da
Ciência e Tecnologia, the Deutsche Forschungsgemeinschaft and the
Collaborating Institutions in the Dark Energy Survey. The
Collaborating Institutions are Argonne National Laboratories, the
University of California at Santa Cruz, the University of Cambridge,
Centro de Investigaciones Energeticas, Medioambientales y
Tecnologicas-Madrid, the University of Chicago, University College
London, the DES-Brazil Consortium, the Eidgenoessische Technische
Hochschule (ETH) Zurich, Fermi National Accelerator Laboratory, the
University of Edinburgh, the University of Illinois at
Urbana-Champaign, the Institut de Ciencies de l'Espai (IEEC/CSIC), the
Institut de Fisica d'Altes Energies, the Lawrence Berkeley National
Laboratory, the Ludwig-Maximilians Universität and the associated
Excellence Cluster Universe, the University of Michigan, the National
Optical Astronomy Observatory, the University of Nottingham, the Ohio
State University, the University of Pennsylvania, the University of
Portsmouth, SLAC National Laboratory, Stanford University, the
University of Sussex, and Texas A\&M University.

This research made use of Astropy, a community-developed core Python package
for Astronomy \citep{astropy}, IPython software \citep{perez07} and 
Matplotlib library \citep{hunter}.

The research leading to these results has received funding from the
European Research Council under the European Union's Seventh Framework
Programme (FP/2007-2013)/ERC Grant Agreement no. 308024.

\end{document}

%% file: table.tex
\tabletypesize{\scriptsize}
\begin{deluxetable*}{ccccccccccccc}
\tablecaption{Parameters of the discovered MW satellites.}
\tablehead{
\colhead{Name} &
\colhead{$\alpha$} &
\colhead{$\delta$} &
\colhead{Signif\tablenotemark{a}} &
\colhead{m$-$M\tablenotemark{b}} &
\colhead{Dist$_\odot$\tablenotemark{b}} &
\colhead{M$_V$} &
\colhead{$r_{\rm maj}$} &
\colhead{r$_{1/2}$\tablenotemark{c}} &
\colhead{r$_{1/2}$\tablenotemark{d}} &
\colhead{ellipticity} &
\colhead{PA} &
\colhead{BF\tablenotemark{e}}\\
\colhead{ } & \colhead{[deg]} & \colhead{[deg]} & \colhead{icance} & \colhead{[mag]} & \colhead{[kpc]} & \colhead{[mag]} & \colhead{[arcmin]} & \colhead{[arcmin]} & \colhead{[pc]} & \colhead{} & \colhead{[deg]} & \colhead{}\
}

\startdata
Reticulum 2 & $53.9256$ & $-54.0492$ & 48.5 & 17.4 & 30 & -2.7$\pm$0.1 & $3.37^{+0.23}_{-0.13}$ & $3.64^{+0.21}_{-0.12}$ & $32^{+1.9}_{-1.1}$ & $0.59^{+0.02}_{-0.03}$ & 71$\pm$1 & $>$1000\\
Eridanus 2 & $56.0878$ & $-43.5332$ & 31.5 & 22.9 & 380 & -6.6$\pm$0.1 & $1.19^{+0.12}_{-0.10}$ & $1.53^{+0.14}_{-0.09}$ & $169^{+16.0}_{-9.8}$ & $0.40^{+0.06}_{-0.08}$ & 81$\pm$6 & 1113\\
Horologium 1 & $43.8820$ & $-54.1188$ & 28.4 & 19.5 & 79 & -3.4$\pm$0.1 & $0.84^{+0.17}_{-0.08}$ & $1.31^{+0.19}_{-0.14}$ & $30^{+4.4}_{-3.3}$ & $<0.28$ & \nodata & 0.35\\
Pictoris 1 & $70.9475$ & $-50.2830$ & 17.3 & 20.3 & 114 & -3.1$\pm$0.3 & $0.70^{+0.23}_{-0.13}$ & $0.88^{+0.27}_{-0.13}$ & $29^{+9.1}_{-4.4}$ & $0.47^{+0.12}_{-0.29}$ & 78$\pm$23 & 1.41\\
Phoenix 2 & $354.9975$ & $-54.4060$ & 13.9 & 19.6 & 83 & -2.8$\pm$0.2 & $0.82^{+0.27}_{-0.12}$ & $1.09^{+0.26}_{-0.16}$ & $26^{+6.2}_{-3.9}$ & $0.47^{+0.08}_{-0.29}$ & 164$\pm$54 & 1.81\\
Indus 1 & $317.2044$ & $-51.1656$ & 13.7 & 20.0 & 100 & -3.5$\pm$0.2 & $0.84^{+0.35}_{-0.17}$ & $1.26^{+0.45}_{-0.27}$ & $37^{+13.1}_{-8.0}$ & $<0.38$ & \nodata & 0.46\\
Grus 1\tablenotemark{f} & $344.1765$ & $-50.1633$ & 10.1 & 20.4 & 120 & -3.4$\pm$0.3 & $1.33^{+0.74}_{-0.26}$ & $1.77^{+0.85}_{-0.39}$ & $62^{+29.8}_{-13.6}$ & $0.41^{+0.20}_{-0.28}$ & 4$\pm$60 & 1.01\\
Eridanus 3 & $35.6897$ & $-52.2837$ & 10.1 & 19.7 & 87 & -2.0$\pm$0.3 & $0.38^{+0.43}_{-0.05}$ & $0.54^{+0.50}_{-0.10}$ & $14^{+12.5}_{-2.6}$ & $0.27^{+0.30}_{-0.16}$ & 83$\pm$36 & 0.81\\
Tucana 2 & $342.9796$ & $-58.5689$ & 8.3 & 18.8 & 57 & -3.8$\pm$0.1 & $7.67^{+1.02}_{-1.18}$ & $9.83^{+1.66}_{-1.11}$ & $165^{+27.8}_{-18.5}$ & $0.39^{+0.10}_{-0.20}$ & 107$\pm$18 & 2.14\\
\enddata
\tablenotetext{a}{Significance of detection in $\sigma$ (t-value)}
\tablenotetext{b}{The uncertainty in distance modulus is estimated to be 0.1$-$0.2 mag}
\tablenotetext{c}{Half-light radius takes into account the ellipticity of the object via $\sqrt{1-e}$ multiplier}
\tablenotetext{d}{The error on distance was not propagated to the physical size}
\tablenotetext{e}{Bayes factor for the elliptical vs circular model}
\tablenotetext{f}{As this object is located very close to the CCD chip gap, its morphological properties should be treated with caution}
\tablecomments{Morphological parameters of the satellites are maximum a posteriori estimates with 68\% (1$\sigma$) credible intervals, or limits}
\label{tab:params}\end{deluxetable*}